\documentclass[sigconf]{acmart}

\copyrightyear{2026}
\acmYear{2026}
\setcopyright{cc}
\setcctype{by}

\acmConference[RecSys '26]
  {20th ACM Conference on Recommender Systems}
  {September 27--October 02, 2026}
  {Minneapolis, MN, USA}

\acmBooktitle{20th ACM Conference on Recommender Systems
  (RecSys '26), September 27--October 02, 2026,
  Minneapolis, MN, USA}

\acmDOI{10.1145/3773078.3831754}
\acmISBN{979-8-4007-2284-4/2026/09}

\usepackage{amsthm}
\usepackage{bibunits}
\usepackage{algorithm}
\usepackage{algorithmic}
\usepackage{cleveref}
\usepackage{enumitem}
\usepackage{nicefrac}
\usepackage{multirow}
\usepackage{url}

\def\I{{\bf I}}

\def\R{{\bf R}}

\def\s{{\bf s}}

\def\x{{\bf x}}

\def\U{{\bf U}}
\def\u{{\bf u}}
\def\V{{\bf V}}
\def\v{{\bf v}}
\def\W{{\bf W}}
\def\w{{\bf w}}
\def\0{{\bf 0}}
\def\1{{\bf 1}}

\def\NM{{\mathcal N}}

\def\LM{{\mathcal L}}

\def\RB{{\mathbb R}}
\def\EB{{\mathbb E}}

\renewcommand{\tilde}{\widetilde}
\renewcommand{\hat}{\widehat}

\def\muu{\mbox{\boldmath$\mu$\unboldmath}}
\def\Si{\mbox{\boldmath$\Sigma$\unboldmath}}

\def\argmax{\mathop{\rm argmax}}

\def\cite#1{\citep{#1}}

\numberwithin{theorem}{section}
\numberwithin{lemma}{section}
\numberwithin{remark}{section}
\numberwithin{cor}{section}

\usepackage{amsmath,amsfonts,bm}

\def\figref#1{Fig.~\ref{#1}}
\def\tabref#1{Table~\ref{#1}}

\def\eqref#1{Eqn.~\ref{#1}}
\def\eqnref#1{Eqn.~\ref{#1}}

\def\algref#1{Algorithm~\ref{#1}}

\def\1{\bm{1}}

\DeclareMathAlphabet{\mathsfit}{\encodingdefault}{\sfdefault}{m}{sl}
\SetMathAlphabet{\mathsfit}{bold}{\encodingdefault}{\sfdefault}{bx}{n}

\def\blue#1{\textcolor{black}{#1}}

\defaultbibliographystyle{ACM-Reference-Format}
\defaultbibliography{main}

\begin{document}
\title{Probabilistic Residual Learning for Online Recommendations}





\author{Wenyuan Wang}
\authornote{Co-first authors.}
\affiliation{%
  \institution{Rutgers University}
  \city{Piscataway}
  \state{New Jersey}
  \country{United States}
}
\email{ww462@scarletmail.rutgers.edu}

\author{Yusong Zhao}
\authornotemark[1]
\authornote{Corresponding authors.}
\affiliation{%
  \institution{Rutgers University}
  \city{Piscataway}
  \state{New Jersey}
  \country{United States}
}
\email{yz1635@scarletmail.rutgers.edu}

\author{Zihao Xu}
\authornotemark[1]
\affiliation{%
  \institution{Rutgers University}
  \city{Piscataway}
  \state{New Jersey}
  \country{United States}
}
\email{zihao.xu@rutgers.edu}

\author{Hengyi Wang}
\authornotemark[1]
\affiliation{%
  \institution{Rutgers University}
  \city{Piscataway}
  \state{New Jersey}
  \country{United States}
}
\email{hw514@scarletmail.rutgers.edu}

\author{Qi Xu}
\authornotemark[1]
\affiliation{%
  \institution{Meta}
  \city{Sunnyvale}
  \state{California}
  \country{United States}
}
\email{xuqi0511@gmail.com}

\author{Zhigang Hua}
\authornotemark[1]
\authornotemark[2]
\affiliation{%
  \institution{Meta}
  \city{Sunnyvale} 
  \state{California}
  \country{United States}
}
\email{zhua@meta.com}

\author{Yan Xie}
\affiliation{%
  \institution{Meta}
  \city{Sunnyvale} 
  \state{California}
  \country{United States}
}
\email{yanxie@meta.com}

\author{Yi Wang}
\affiliation{%
  \institution{Rutgers University}
  \city{Piscataway}
  \state{New Jersey}
  \country{United States}
}
\email{yw1013@scarletmail.rutgers.edu}

\author{Zihao Zhao}
\affiliation{%
  \institution{Rutgers University}
  \city{Piscataway}
  \state{New Jersey}
  \country{United States}
}
\email{zz1009@rutgers.edu}

\author{Bo Long}
\affiliation{%
  \institution{Meta}
  \city{Sunnyvale} 
  \state{California}
  \country{United States}
}
\email{bolong@meta.com}

\author{Chengzhi Mao}
\affiliation{%
  \institution{Rutgers University}
  \city{Piscataway}
  \state{New Jersey}
  \country{United States}
}
\email{chengzhi.mao@rutgers.edu}

\author{Shuang Yang}
\affiliation{%
  \institution{Meta}
  \city{Sunnyvale} 
  \state{California}
  \country{United States}
}
\email{shuangyang@meta.com}

\author{Hengguan Huang}
\authornotemark[2]
\affiliation{%
  \institution{University of Copenhagen}
  \city{Copenhagen}
  \country{Denmark}
}
\email{hengguan.huang@sund.ku.dk}

\author{Hao Wang}
\authornotemark[2]
\affiliation{%
  \institution{Rutgers University \& UIUC}
  \city{Piscataway}
  \state{New Jersey}
  \country{United States}
}
\email{hwangml@illinois.edu}

\renewcommand{\shortauthors}{Wang et al.}


\newcommand{\fix}{\marginpar{FIX}}
\newcommand{\new}{\marginpar{NEW}}

\begin{abstract}
Modern recommender systems are typically based on deep learning (DL) models, where a dense encoder learns representations of users and items. As a result, these systems often suffer from the black-box nature and computational complexity of the underlying models, making it difficult to systematically enhance their recommendation capabilities. To address this problem, we propose \emph{Probabilistic Residual Learning (PRL)}, a causal Bayesian recommendation model that models the residual between ground-truth and base predictions, enabling targeted refinement of existing systems. 
{Specifically, PRL (1) probabilistically groups users for localized residual modeling, (2) models domain-level confounders that influence user and item representations, and (3) aggregates cluster-specific residual predictions over the confounders using do-calculus.}
Experiments demonstrate that our \emph{plug-and-play} PRL is compatible with various base DL recommender systems, improving their performance while automatically discovering meaningful user clusters.
\end{abstract}


\maketitle

\begin{figure*}[t!]
    \centering
    \includegraphics[width=0.8\textwidth]{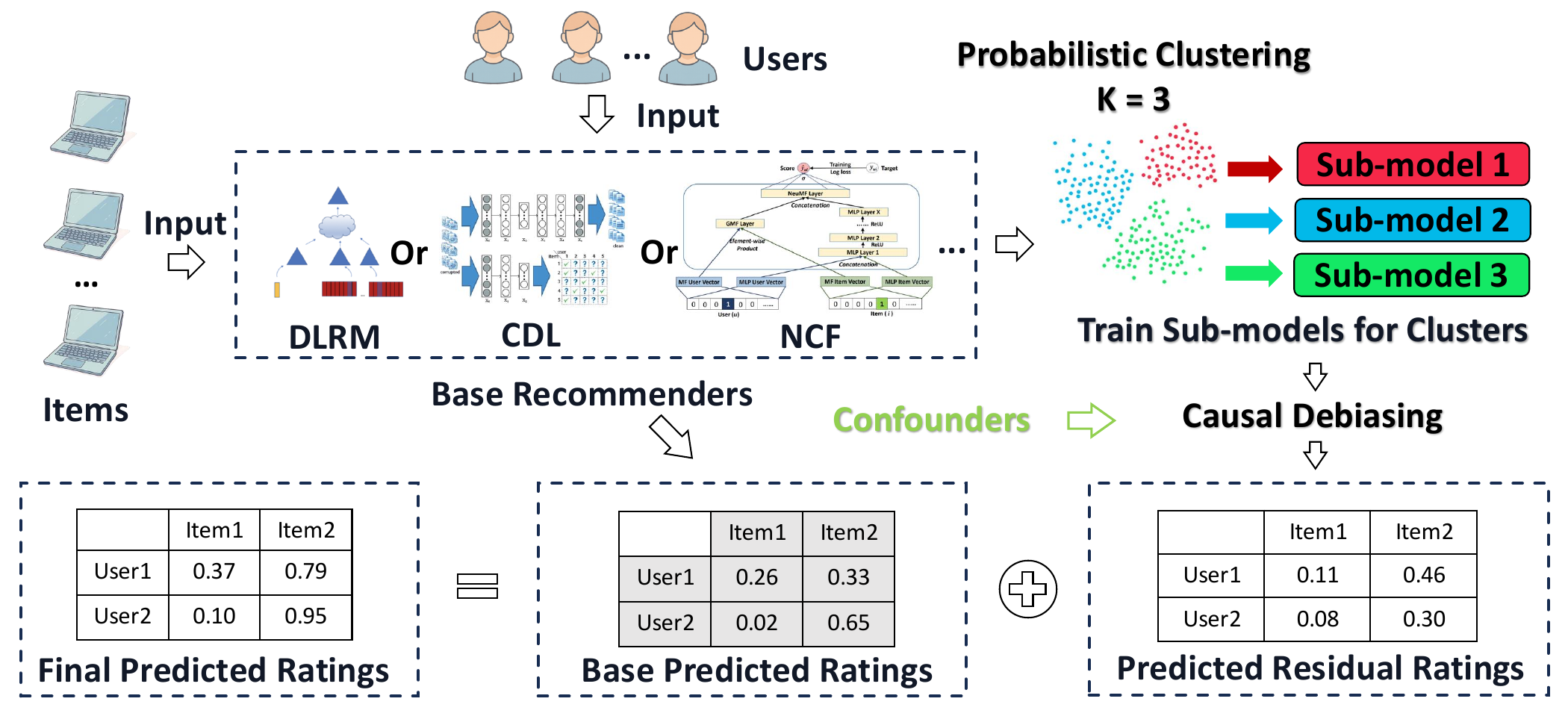} 
    \caption{{\textbf{Overview of PRL.} During training, PRL models the
    residuals of a fixed base recommender, probabilistically clusters users,
    and learns cluster-specific correction models. During inference, it
    predicts a confounder-adjusted residual using the inferred cluster and
    adds it to the base prediction.}}
    \Description{Overview of PRL. A fixed base recommender first produces
    base rating predictions. PRL computes residual ratings, probabilistically
    groups users into clusters, and learns cluster-specific residual models
    conditioned on domain factors. At inference time, the predicted residual
    is added to the base prediction to obtain the final rating.}
    \label{fig:teaser} 
    
\end{figure*}
\section{Introduction}

Over the past decade, personalized recommendations have significantly improved user experiences in domains such as e-commerce and social media. The recommender systems driving these advancements often rely on sophisticated deep learning (DL) models~\cite{ChungGCB14,vaswani2017attention,srgnn} capable of handling vast amounts of data, enabling highly accurate predictions and personalized interactions. {Despite their effectiveness, these models are often costly or difficult to systematically adapt to new domains, particularly under cold-start domain shifts such as changes in user markets.} Cold-start scenarios, a critical problem in recommendation systems, exacerbate these issues due to the presence of heterogeneous features and the influence of diverse and spurious patterns. As a result, existing models exhibit notably low performance in such settings.

{Existing work~\cite{DBLP:conf/sigir/Yuan0KZ20,DBLP:conf/sigir/WuYCLH020,
DBLP:conf/sigir/BiSYWWX20a,li2019zero,DBLP:conf/sigir/Hansen0SAL20,
DBLP:conf/sigir/LiangXYY20,DBLP:conf/sigir/ZhuSSC20,
liu2020heterogeneous,kweon2024perk}
often addresses domain shift through shared users or items.
However, such overlap is often unavailable in practice; for example,
users and items may be disjoint across countries. This setting also
requires modeling shared confounders, as exposure and popularity
effects may induce domain-specific correlations that transfer poorly.
Additionally, existing methods often overlook latent user clusters,
missing opportunities for user-cluster-targeted enhancement.}


To address these problems, we 
draw inspiration from the hierarchical Bayesian DL framework~\cite{BDL,BDLSurvey} to 
propose causal hierarchical Bayesian DL model, dubbed \emph{Probabilistic Residual Learning (PRL)}, as a \emph{plug-and-play} framework to improve and potentially interpret any base recommender systems \blue{in cross-domain settings}.
\figref{fig:teaser} shows the simplified overview of our framework. 
\textbf{During the training stage}, given any base recommender (e.g., DLRM~\cite{DLRM}, CDL~\cite{cdl}, or NCF~\cite{NCF}), PRL is learned by jointly (1) computing its predicted ratings and the residual ratings (i.e., the difference between the ground-truth ratings and the base model's predicted ratings), (2) dividing users into latent clusters based on the residual ratings using our probabilistic clustering method, (3) training a sub-model for each latent user cluster. \textbf{During the inference stage}, once PRL is learned, given a new user-item pair, PRL can then (1) estimate the user's cluster ID to select the proper sub-model, (2) use this sub-model to perform causal inference to debias potential confounders and predict the residual rating, and (3) add this predicted residual rating to the base predicted rating to obtain the final predicted rating, thereby producing the final recommendation. Notably, PRL is \emph{plug-and-play}, i.e., it is compatible with any base DL recommendation model and can enhance the original model's performance. {Our contributions are as follows:}

\begin{itemize}[nosep]
    \item {We formulate cross-domain adaptation as a plug-and-play
    probabilistic residual correction problem for fixed base recommenders.}
    
    \item {We develop a hierarchical Bayesian model that combines
    probabilistic user clustering with causal adjustment of domain
    confounders.}
    
    \item {Experiments across datasets and base recommenders
    demonstrate improved cold-start cross-domain recommendation.}
\end{itemize}

\section{{Probabilistic Residual Learning}}\label{sec:prl}

\subsection{Problem Setting and Notations}
Consider a recommendation dataset containing $I$ users and $J$ items. A DL encoder $f_v(\cdot): \RB^d\rightarrow \RB^h$ encodes each item $j$'s raw features $\x_j^v \in \RB^d$ into $f_v(\x_j^v)$. 
For a given user $i$ and an item $j$, there is a ground-truth rating $R_{ij}\in \RB$, a base predicted rating $\hat{R}_{ij}\in \RB$ provided by a base recommender, and a residual rating $\tilde R_{ij} = R_{ij} - \hat{R}_{ij}$. There is a latent cluster ID $k$ ($k \in \{1,...,K\}$) that indicates which user group user $i$ belongs to. We assume that there exists a user latent vector $\u_i\in \RB^h$ for each user $i$ and an item latent vector $\v_j\in \RB^h$ for each item $j$; 
they are both impacted by a causal confounder $\s \in \RB^{g}$, where $g\ll h$. 

Our goal is to predict the final rating $R$ using the {residual $\tilde{R}$}, i.e., $R = \hat{R} + \tilde{R}$, where $\hat{R}$ represents the rating from the original (base) DL recommender. When the original recommender is provided, $\hat{R}$ is fixed; therefore we only need to learn $\tilde{R}$ in order to predict the final rating $R$. For generality, we assume $M$ domains, {where $m_i$ and $m_j$ denote} the domain ID of user $i$ and item $j$, respectively.

\begin{figure}[t]
    \centering
\includegraphics[width=0.8\linewidth]{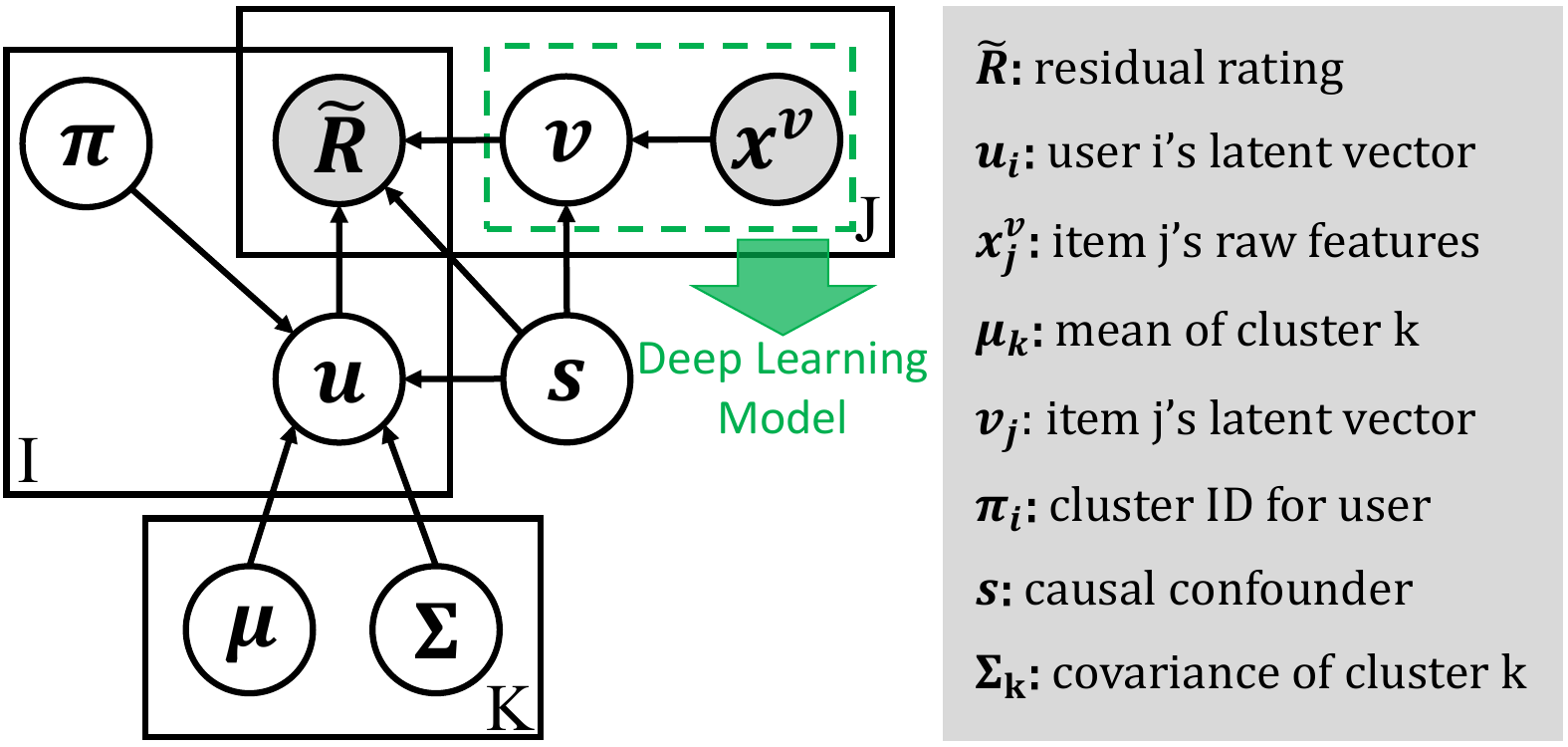}
\caption{\label{fig:pgm}
{Probabilistic graphical model of PRL. The cluster variable
$\pi_i$ selects the user prior, while the domain factor $\s_m$
influences $\u_i$, $\v_j$, and the residual rating $\tilde R_{ij}$.}}
\Description{The probabilistic graphical model of PRL. The user cluster
variable pi selects a cluster-specific prior for the user latent variable.
The domain factor influences the user latent variable, item latent
variable, and residual rating. Item features are encoded by a deep
learning model to inform the item latent variable.}
\vskip -0.5cm 
\end{figure}

\subsection{Method Overview}

{PRL jointly models cluster-specific users, content-informed
items, and confounder-adjusted residuals. \figref{fig:pgm} shows their
dependencies and variational formulation.}

\textbf{Generative Process.}
Below we describe the generative process of PRL shown in~\figref{fig:pgm}.

{
For each domain $m$, PRL draws a confounder
$\s_m\sim\NM(\0,\I)$. For each item $j$, it draws
\begin{align}
\v_j \sim p(\v_j\mid\x_j^v,\s_m)
=
PoG\!\left(
f_v(\x_j^v),\W^v\s_m,
\Lambda_v^{-1}\I,\lambda_v^{-1}\I
\right).
\label{eq:pog}
\end{align}
For each user $i$, PRL draws a cluster assignment
$\pi_i\sim p(\pi_i\mid\theta)$ and
\[
\u_i\sim
\NM(\muu_{\pi_i}+\W^u\s_m,\Si_{\pi_i}).
\]
Finally, it draws the residual rating
\[
\tilde R_{ij}\sim
\NM\!\left(
\u_i^\top\v_j+\w^{R\top}\s_m,
\lambda_{\tilde R_{ij}}^{-1}
\right).
\]
Here, $\W^u$, $\W^v$, and $\w^R$ are globally shared parameters.}

{The product-of-Gaussians prior in~\eqnref{eq:pog} combines the
content-based estimate $f_v(\x_j^v)$ and the domain-conditioned
estimate $\W^v\s_m$. The resulting distribution is also
Gaussian~\cite{pog}, with
\begin{align}
\muu_{pog}
&=
\frac{\Lambda_v f_v(\x_j^v)+\lambda_v\W^v\s_m}
{\Lambda_v+\lambda_v},
&
\lambda_{pog}
&=
\Lambda_v+\lambda_v.
\label{eq:pog_para}
\end{align}}

\textbf{Model Factorization}.
As shown in~\figref{fig:pgm}, we factorize the generative model into four conditional distributions: 
\begin{align}
    &p(\u_i, \v_j, \pi, \tilde R_{ij}|\{\muu_k,\Si_k\}_{k=1}^K, \x_j^v, \s_m) \nonumber \\
    =  ~&p(\tilde R_{ij}|\u_i, \v_j, \s_m)   p(\u_i|\{\muu_k,\Si_k\}_{k=1}^K,\s_m, \pi)   p(\v_j|\x_j^v,\s_m) p(\pi). \label{eq:p_factor} 
\end{align}
$p(\pi|\theta)$ is the prior distribution for $\pi$. Each of the remaining distributions is assumed as a Gaussian distribution and is shown as follows:
%
\begin{align}
p(\tilde R_{ij}|\u_{i}, \v_j, \s_m)&=\NM(\u_{i}^\top\v_j+\blue{\w^R}^\top \s_{m}, \lambda_{\tilde R_{ij}}^{-1}), \label{eq:p_R_ij}\\
p(\u_i|\{\muu_k,\Si_k\}_{k=1}^K, \s_m, \pi) &= \NM(\muu_{\pi_i}+\blue{\W^u} \s_m , \Si_{\pi_i}), \label{eq:infer_u}\\
p(\v_j|\x_j^v, \s) &= 
        PoG(f_v(\x_j^v),\W^v \s_{m},\Lambda_v^{-1}\I,\lambda_v^{-1}\I), 
\end{align}
%
where $i$ and $j$ {refer} to the user index and the item index, respectively. We employ variational distributions $q(\u_i, \v_j|\x_j^v)$ to approximate the posterior distributions of $\u_i$ and $\v_j$. 
\begin{align}
q(\{\u_i\}_{i=1}^I, \{\v_j\}_{j=1}^J) = \prod\nolimits_{i=1}^I q(\u_i) \prod\nolimits_{j=1}^J q(\v_j). \label{eq:q_factor}
\end{align}
More specifically, {we assume} $q(\v_j)$ follows a {gaussian distribution}:
\begin{align}
q(\v_j) &=\NM(\muu_{\v_j}, \Lambda_v^{-1}\I)\label{eq:infer_v}. 
\end{align}
Here, $j$ is the item index, $\Lambda_v \in \RB$ refers to the precision. \blue{Similarly:
\begin{align}
q(\u_i) &=\NM(\muu_{\u_i}, \Lambda_u^{-1}\I),    
\end{align}
where $i$ is the user index, and $\Lambda_u \in \RB$ is the precision. Different users and items have different approximate posteriors. We also use a categorical variational distribution $q(\pi)$ to approximate the posterior distribution of $\pi$ (more details below).}

\textbf{Learning Objective.} We maximize an evidence lower bound (ELBO) as our learning objective for both {the generative and inference models}.  
\begin{align}\label{eq:evidence}
    &\LM_{ELBO}(\x^v_j, \tilde R_{ij}) \nonumber\\
    =~& \EB_{q(\u_i) q(\v_j)}\big[\log p(\u_i, \v_j,\tilde R_{ij}|\{\muu_k,\Si_k\}_{k=1}^K, \x^v_j, \s_{m}, \pi)\big] \nonumber \\
    &+\EB_{q(\pi)}[p(\pi|\theta)] - \EB_{q(\pi)}[q(\pi)]\nonumber \\
    &-\EB_{q(\u_i) q(\v_j)}\big[\log q(\v_j)] \nonumber \\
    &-\EB_{q(\u_i) q(\v_j)}\big[\log q(\u_i)]. 
\end{align}

%
Combining \eqref{eq:p_factor} and \eqref{eq:q_factor}, we obtain the following decomposition:
\begin{align}
    &\LM_{ELBO}(\x^v_j, \tilde R_{ij}) \nonumber\\
    =~&\blue{- D_{KL}\big(q(\u_i) \Vert p(\u_i|\{\muu_k,\Si_k\}_{k=1}^K, \s_m, \pi)\big)}\label{eq:e_mu_sigma}\\
    &+ \EB_{q(\u_i) q(\v_j) q(\pi)}\big[\log p(\tilde R_{ij}|\u_i,\v_j, \s_m, \pi)\big] \label{eq:r_res}\\
    &- D_{KL}\big(q(\v_j) \Vert p(\v_j|\x_j^v,\s_m)\big) \label{eq:kl}, \\
    &- D_{KL}\big(q(\pi) \Vert p(\pi | \theta)\big) \label{eq:kl_pi}, 
\end{align}

where $D_{KL}(\cdot \Vert \cdot)$ is the Kullback-Leibler (KL) divergence. For ~\eqref{eq:e_mu_sigma}, we compute the log likelihood for each cluster $k$ as
\begingroup\makeatletter\def\f@size{9.7}\check@mathfonts
\def\maketag@@@#1{\hbox{\m@th\normalsize\normalfont#1}}
\begin{align}\label{eq:e_mu_sigma_expand}
    &\log p(\{\u_i\}_{i \in I_k}\mid \{\muu_k,\Si_k\}, \s_m, \pi) \nonumber\\
    =~&  -\frac{1}{2} \sum_{i \in I_k} [ \log |\Si_k| \nonumber + (\u_i - \muu_k - \blue{\W^u} \s_m)^\top \Si_k^{-1} (\u_i - \muu_k - \W^u \s_m) ] \nonumber\\
    &\quad\quad\quad  +C,
\end{align}
\endgroup
where $i$ is the user index, $I_k$ is the set of user index that belongs to cluster $k$, and $C$ is a constant.

Similarly, all the other terms can be expanded as:
%
\begin{align}
&\log p(\tilde R_{ij}|\u_i,\v_j,\s) = - \frac{\lambda_{\tilde R_{ij}}}{2}  \left( \tilde R_{ij} - \u_i^\top \v_j - \blue{\w^R}^\top \s_m  \right)^2 + C,\\
&D_{KL}\big(q(\v_j) \Vert p(\v_j|\x_j^v,\s_m)\big) 
= -\frac{\lambda_v}{2} \| \blue{\muu_{\v_j}} - \blue{\W^v} \s_m \|^2 \nonumber \\
&\quad\quad\quad- \frac{\Lambda_v}{2} \| \blue{\muu_{\v_j}} - f_v(\mathbf{x}_j^v) \|^2 + C . 
\end{align}
\blue{Here $C$ can be omitted because we treat scalars like $\lambda_v \in \mathbb{R}$, $\lambda_u \in \mathbb{R}$, $\Lambda_v \in \mathbb{R}$, and $\Lambda_u \in \mathbb{R}$ as constants.}

\textbf{Intuition for Each Term in~\eqnref{eq:evidence}.} 
Below, we describe the intuition of each term in~\eqnref{eq:evidence}:
\begin{enumerate}
    \item \textbf{Regularize Latent Variable $\u_i$~(\eqref{eq:e_mu_sigma}).} 
    The KL term $D_{KL}\big(q(\u_i) \Vert p(\u_i|\{\muu_k,\Si_k\}_{k=1}^K, \s_m, \pi)\big)$
    aims to regularize user $i$'s latent variable $\u_i$, ensuring $\u_i$ is close to the center of its corresponding user cluster $\pi_i$, and therefore close to other users' latent embeddings in the same cluster.
     \item \textbf{Predict Residual Rating $\tilde R_{ij}$ from $\u_i$ and $\v_j$~(\eqref{eq:r_res}).} $p(\tilde R_{ij}|\u_i,\v_j, \s_m)$ {uses} the inferred $\u_i$, $\v_j$, and the causal confounder $\s_m$ to predict the residual rating, thereby encouraging $\u_i$ and $\v_j$ to retain more information to maximize prediction performance. 
   \item \textbf{Regularize Latent Variable $\v_j$~(\eqref{eq:kl}).} Similar to~\eqref{eq:e_mu_sigma}, $D_{KL}(q(\v_j) \Vert p(\v_j|\x^v_j,\s_m))$ is the KL divergence term between the inference model $q(\cdot|\x^v_j)$ and the generative model $p(\cdot|\s_m)$; this encourages the inferred posterior $q(\v_j|\x^v_j)$ to be close to the prior distribution $p(\v_j|\s_m)$.
   \item \blue{\textbf{Regularize Latent Variable $\pi$~(\eqref{eq:kl_pi}).}   $D_{KL}\big(q(\pi) \Vert p(\pi | \theta)\big)$ is the KL divergence term between the categorical variational distribution $q(\pi)$ and the prior $p(\pi | \theta)$; this encourages the inferred posterior $q(\pi)$ to be close to the prior $p(\pi | \theta)$.}
\end{enumerate}
    

\subsection{Inference and Learning}

In our framework, we need to learn several parameters, including the Gaussian parameters $\{\muu_k, \Si_k\}_{k=1}^K$, user latent $\mathbf{u}$, item latent $\mathbf{v}$, and the parameters of the functions $f_x(\cdot)$ and $f_v(\cdot)$, as well as
\blue{$\mathbf{W}^u$, $\mathbf{W}^v$, and $\mathbf{w}^R$}.
The following sections detail the learning process for all these parameters. The complete algorithm is outlined in \algref{alg:learn}.

\textbf{1) $\{\muu_k, \Si_k\}_{k=1}^K$}. To optimize $\{\muu_k, \Si_k\}_{k=1}^K$, we take {derivatives} of~\eqref{eq:e_mu_sigma_expand} w.r.t. $\muu_k$ and $\Si_k$ as follows:
\begin{align}\label{eq:grad_mu}
    &\frac{\partial \LM}{\partial \muu_k} = \Si_k^{-1} \left( \u_i - \muu_k - \blue{\W^u} \s_m \right), \\
    \label{eq:grad_sigma}
    &\frac{\partial \LM}{\partial \Si_k} = \frac{1}{2} \Si_k^{-1} \left[ \left( \u_i - \muu_k - \blue{\W^u} \s_m \right) \left( \u_i - \muu_k - \blue{\W^u} \s_m \right)^\top 
    - \Si_k \right] \Si_k^{-1}.
\end{align}
Setting~\eqref{eq:grad_mu} and~\eqref{eq:grad_sigma} to zero leads to the following update rules, respectively:
\begin{align}\label{eq:update_mu}
    &\muu_k = \frac{1}{\vert I_k\vert} \sum_{i \in I_k} \left( \u_i - \blue{\W^u} \s_m \right),\\
    \label{eq:update_sigma}
    &\Si_k = \frac{1}{\vert I_k\vert} \sum_{i \in I_k} \left( \u_i - \muu_k - \blue{\W^u} \s_m \right) \left( \u_i - \muu_k - \blue{\W^u} \s_m \right)^\top,
\end{align}
where $\I_k$ is the set of user index $i$ that belongs to cluster $k$.

\textbf{2) $\u_i, \v_j$, and $\pi_i$}. 
After computing the gradients of \eqref{eq:evidence}  w.r.t. \blue{the means of $\u_i \sim \NM(\muu_{\u_i}, \Lambda_u^{-1}\I)$ (i.e., $\muu_{\u_i}$) and $\v_j \sim \NM(\muu_{\v_j}, \Lambda_v^{-1}\I)$ (i.e., $\muu_{\v_j}$)}, we obtain the following update rules:
\begin{align}
\muu_{\u_i}
={}&
(\Si_{\pi_i}\V \lambda_{\tilde R_{(i,:)}}\V^\top + \I)^{-1}
\Big[
\muu_{\pi_i} + \blue{\W^u}\s_m
\notag\\
&\qquad\qquad
+ \Si_{\pi_i}\V \lambda_{\tilde R_{(i,:)}}
(\tilde \R_{(i,:)} - \blue{\w^R}^\top \s_m \I)
\Big],
\label{eq:update_u}
\\[2pt]
\muu_{\v_j}
={}&
\Big[
\U \lambda_{\tilde R_{(:,j)}}\U^\top
+ (\lambda_v - \Lambda_v)\I
\Big]^{-1}
\notag\\
&\qquad\cdot
\Big[
\lambda_v \blue{\W^v}\s_m
- \Lambda_v f_v(\x_j^v)
\notag\\
&\qquad\qquad
+ \U \lambda_{\tilde R_{(:,j)}}
(\tilde \R_{(:,j)} - \blue{\w^R}^\top \s_m \I)
\Big].
\label{eq:update_v}
\end{align}
{PRL is efficient; \eqref{eq:update_u} uses only the user's rated items, so its cost
scales with rated items rather than the full catalog.}

Note that here $\U$ and $\V$ refer to user latent matrix $(\u_i)_{i=1}^I$ and item latent matrix $(\v_j)_{j=1}^J$. $\tilde \R_{(i,:)} := (\tilde R_{i1}, \cdots, \tilde R_{iJ})^\top$,  $\tilde \R_{(:,j)} := (\tilde R_{1j}, \cdots, \tilde R_{Ij})^\top$ . $\lambda_{\tilde R_{(i,:)}} := \text{diag}(\lambda_{\tilde R_{i1}}, \cdots, \lambda_{\tilde R_{iJ}})$, $\lambda_{\tilde R_{(:,j)}} := \text{diag}(\lambda_{\tilde R_{1j}}, \cdots, \lambda_{\tilde R_{Ij
}})$.

\blue{With a uniform prior distribution on $\pi_i$, i.e., $p(\pi_i=k|\theta)=\frac{1}{K}$, the estimation of $\pi_i$ is similar to that of a Gaussian mixture model (GMM). Specifically, we can approximate $p(\pi_i = k | u_i, v_j, x_j^v, \{\mu_k, \Sigma_k\}^K_{k=1})$ using 
\begin{align}
q(\pi_i = k) =\frac{\mathcal{N}(\u_i;\mu_k + W_u s_{m_i},\Sigma_k)}
{\sum_{l=1}^K \mathcal{N}(\u_i;\mu_l + W_u s_{m_i},\Sigma_l)}, \label{eq:update_pi}
\end{align}
which is obtained by maximizing the ELBO \eqnref{eq:e_mu_sigma}$\sim$\ref{eq:kl_pi}.} \blue{We then choose the optimal (most probable) cluster for each user $i$, i.e., the cluster assignment for user $i$ is set to $\pi_i = \argmax_k q(\pi_i = k)$, which is used to compute~\eqnref{eq:update_u}$\sim$\ref{eq:update_v}.}
\begin{algorithm}[!t]
\vskip 0.1cm
\raggedright
 \caption{Learning Algorithm of PRL}\label{alg:learn}
 \textbf{Input:} Raw item features $\x^v$, initialized $f_x(\cdot)$ and $f_v(\cdot)$ parameters, $\blue{\W^u},\blue{\W^v},\blue{\w^R}$, initialized Gaussian parameters $\{\muu_k, \Si_k\}_{k=1}^K$, and the number of epochs~T.
 
\textbf{for} {$t=1:T$} \textbf{do}{

 \quad\textbf{for} {$m=1:M$} \textbf{do}{

\quad\quad Update $\u_i$ and $\v_j$ using \eqref{eq:update_u} and \eqref{eq:update_v} \blue{update the variational distribution of $\pi_i$ using~\eqnref{eq:update_pi}}.

 \quad\quad Update $\blue{\W^u},\blue{\W^v},\blue{\w^R}$ using \eqref{eq:update_w_u}, \eqref{eq:update_W^v} and \eqref{eq:update_w_r}.
 
 \quad\quad Update the parameters of $f_v(\cdot)$  using gradient ascent of $\LM$ in~\eqref{eq:evidence}.
 
 }
 \quad Update $\{\muu_k, \Si_k\}_{k=1}^K$ using~\eqref{eq:update_mu} and~\eqref{eq:update_sigma}, respectively. 

\textbf{Output:} $f_v(\cdot)$ parameters, $\blue{\W^u},\blue{\W^v},\blue{\w^R}$, and Gaussian parameters $\{\muu_k, \Si_k\}_{k=1}^K$.
 }
\end{algorithm}

\textbf{3) $\blue{\W^u}$, $\blue{\W^v}$, $\blue{\w^R}$.} The update rules for $\blue{\W^u}$, $\blue{\W^v}$, and $\blue{\w^R}$ are as follows:
\begin{align}
    &\blue{\W^u} = \frac{1}{I} (\sum_{i=1}^I \u_i - \sum_{k=1}^K \vert I_k \vert \muu_k )\s_m^\top (\s_m \s_m^\top)^{-1} ,\label{eq:update_w_u} \\
    &\blue{\W^v} = \frac{1}{J} \sum_{j=1}^J \v_j \s_m^\top (\s_m \s_m^\top)^{-1},
    \label{eq:update_W^v}\\ 
    &\blue{\w^R} = \frac{\sum_{i,j} \lambda_{\tilde R_{ij}} (\tilde R_{ij} - \u_i^\top \v_j) }{\sum_{i,j} \lambda_{\tilde R_{ij}}}(\s_m \s_m^\top)^{-1} \s_m \label{eq:update_w_r}.
\end{align}

\textbf{4) Parameters of $f_x(\cdot)$ and $f_v(\cdot)$.} We use gradient ascent of $\LM$ in~\eqref{eq:evidence} to update these parameters. 


\textbf{Inference.} Inference includes the \emph{E-Step} in~\algref{alg:learn}, where PRL \blue{infers latent variables $\u_i$ and $\v_j$}, and updates the parameters of encoder model $f_v(\cdot)$ using gradient ascent of $\LM$ in~\eqref{eq:evidence}. 

\textbf{Learning.} Learning includes the iteration between the \emph{E-Step} and \emph{M-Step} in~\algref{alg:learn} until convergence. In each \emph{M-Step}, we update the Gaussian parameters $\{\muu_k, \Si_k\}_{k=1}^K$ following the update rules from~\eqref{eq:update_mu} and~\eqref{eq:update_sigma}, respectively; \blue{we update learnable parameters $\blue{\W^u},\blue{\W^v}$, $\blue{\w^R}$ following the update rules from~\eqnref{eq:update_w_u},~\eqnref{eq:update_W^v}, and~\eqnref{eq:update_w_r}, respectively.} 
\subsection{Plug-and-Play PRL}
Below we discuss key components of our plug-and-play PRL after learning all parameters with \algref{alg:learn}. 

\textbf{Inferring User Clusters.}
{Given the learned mixture parameters
$\{\muu_k,\Si_k\}_{k=1}^K$, PRL assigns each user to the cluster
$\pi_i$ with the largest posterior probability in~\eqref{eq:update_pi}, similar to a hard EM algorithm.}

\textbf{Isolating Causal Confounders $\s_m$.} 
With the structured causal model (SCM), we estimate a {domain-level confounder representation} $\s_m$ for each domain $m$ by approximating its posterior distribution $p(\s_m|\tilde{R},\x_j^v, \{\muu_k,\Si_k\}_{k=1}^K)$ via variational domain indexing (VDI)~\cite{VDI}. In this way, we can minimize the bias introduced by the causal confounder $\s_m$ when inferring $\u_i$ and $\v_j$ using~\eqref{eq:update_u} and~\eqref{eq:update_v}, respectively. {In practice, $\s_m$ may be instantiated by an observed proxy or by a learned domain index, depending on the dataset. When VDI is used, the domain index $s_m$} can be thought of as the embedding for each domain $m$. For example, in the dataset XMRec where each of the 18 domains contains items and users from one market/country (e.g., France or US), $s_m$ can be thought of as a ``country'' embedding. Interestingly, our preliminary results show that similar countries tend to have similar domain embedding $s_m$ (i.e., domain index)~\cite{VDI} . In other words, $s_m$ captures the similarities among different domains and therefore {provides} valuable information for our recommender systems.

\textbf{Debiasing the Causal Confounders.}
Under our \emph{PRL} framework, for each inferred user cluster $k$, we {perform causal inference} for each user $i$ in {this} cluster to predict the residual $\tilde{R}_{ij}$ (for each item~$j$) while debiasing the causal confounders $\s$. Specifically, with the inferred $\u_i$ and $\v_j$ (using \eqref{eq:update_u} and~\eqref{eq:update_v}) and $\s_m$, we can predict $\tilde{R}_{ij}$ by do-calculus as
    \begin{align}\label{eq:do-calculus}
        p^{(k)}(\tilde{R}_{ij} | do(\u_i), do(\v_j)) 
        = \sum\nolimits_{m=1}^M p^{(k)}(\tilde{R}_{ij} | \u_i, \v_j, \s_m) p(\s_m),
    \end{align}
where $p^{(k)}(\tilde{R}_{ij} | \u_i, \v_j, \s)$ represents the $k$'th sub-model trained from the $k$'th cluster's user data. In practice, we use $k=\pi_i$ ($\pi_i$ is user $i$'s cluster) when predicting user $i$'s rating $\tilde{R}_{ij}$.

\begin{figure}[t]
\centering
\vspace{-0.2cm}
\includegraphics[width=0.5\linewidth]{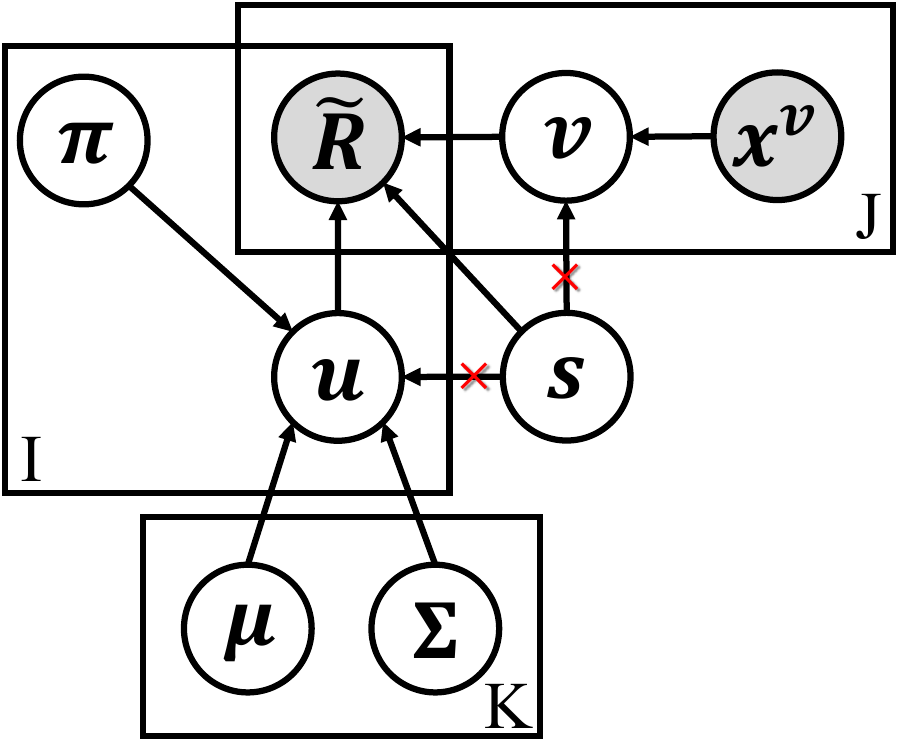}
\vspace{-0.3cm}
\caption{Causal inference in PRL is equivalent to {removing the influence of confounder $\s$ on $\u_i$ and $\v_j$.}}
\Description{The causal graphical model after intervention on the user
and item latent variables. The incoming edges from the domain confounder
$\s$ to $\u_i$ and $\v_j$ are removed, while $\u_i$ and $\v_j$ remain
connected to the residual rating $\tilde R_{ij}$.}
\label{fig:pgm_causal}
\vspace{-0.5cm}
\end{figure}

Note that performing causal inference by intervening $(\u_i,\v_j)$ effectively cuts the relations between the causal confounders $\s$ and $(\u_i,\v_j)$.~\figref{fig:pgm_causal} {demonstrates} the do-calculus that PRL performs for debiasing the causal confounder $\s$. 

\textbf{Intuition behind Do-Calculus: Why Marginalize over the Domain Factor $\s_m$ in~\eqref{eq:do-calculus}.} 
Estimating interventional distributions typically requires actively intervening in a recommender system to collect data, which is often costly and impractical. Instead, PRL leverages do-calculus to estimate the interventional effect $p^{(k)}(\tilde{R}_{ij} \mid do(\u_i), do(\v_j))$ from observational data by marginalizing over the confounder $\s$. This operation reduces reliance on spurious domain-specific correlations. 

In datasets such as XMRec, where each domain corresponds to a country, $\s_m$ can act as a confounder, inducing biases such as exposure or popularity effects. A predictor conditioned on a specific domain may overfit these shortcuts, attributing preferences to country-specific patterns rather than true user-item relationships. For example, a model may associate Bollywood films with cricket gear due to their co-occurrence in a market, not due to intrinsic preference. By marginalizing over $\s_m$ as in~\eqref{eq:do-calculus}, PRL removes such confounding effects and retains residual signals that are more stable and transferable across domains.

\textbf{Summary.} 
To summarize, for each user $i$, PRL causally {infers} the residual rating $\tilde R_i$ as follows:
\begin{enumerate}
    \item Infer the user cluster $\pi_i$ by approximating its posterior \\ $p(\pi_i|\u_i,\v_j, \x^v_j,\{\muu_k,\Si_k\}_{k=1}^K)$.
    \item Infer the residual rating $\tilde{R}_{ij}$ by causal Bayesian model averaging defined in~\eqref{eq:do-calculus}.
    \item Predict the final rating as $R=\tilde{R} + \hat{R}$, where $\hat{R}$ is the base recommender's prediction.
\end{enumerate}

\begin{table*}[!t]
\centering
\vskip -0.3cm
\caption{Performance of PRL with different base models on XMRec. The best results are marked with \textbf{bold face}.\label{tab:avg_country}}
\vskip -0.2cm
\resizebox{0.88\textwidth}{!}{
\setlength{\tabcolsep}{8pt}
\begin{tabular}{c|c|c|c|c|c|c}
\toprule
\textbf{Data} &  \textbf{Method} & \textbf{Recall@20} & \textbf{F1@20} & \textbf{MAP@20} & \textbf{NDCG@20} & \textbf{Precision@20} \\ 
\midrule
\multirow{10}{*}{\shortstack{France, Italy, India $\rightarrow$\\ Japan, Mexico}} 
                        & CDL (Base Model)            & 0.0143 & 0.0016 & 0.0028 & 0.0009 & 0.0009 \\ 
                       & PRL (Full)             & \textbf{0.1091} & \textbf{0.0128} & \textbf{0.0463} & \textbf{0.0108} & \textbf{0.0068} \\ 
\cmidrule{2-7} 
                       & DLRM (Base Model)               & 0.0044 & 0.0004 & 0.0004 & 0.0002 & 0.0002 \\ 
                        & PRL (Full)             &\textbf{ 0.0295} & \textbf{0.0035} & \textbf{0.0048} &\textbf{ 0.0018} & \textbf{0.0018} \\ 
\cmidrule{2-7} 
                       & PerK (Base Model)               & 0.1098 & 0.0128 & 0.0512 & 0.0112 & 0.0068 \\ 
                        & PRL (Full)             & \textbf{0.1635} & \textbf{0.0192 }& \textbf{0.0637} &\textbf{ 0.0151} & \textbf{0.0102} \\ 
\cmidrule{2-7}
                        & NCF (Base Model)               & 0.0131 & \textbf{0.0148} & 0.0026 & 0.0008 & 0.0008 \\ 
                        & PRL (Full)             & \textbf{0.1137} & 0.0137 & \textbf{0.0309} &\textbf{ 0.0090} & \textbf{0.0073} \\    
\cmidrule{2-7}
                        & LightGCN (Base Model)               & 0.0182 & 0.0021 & 0.0050 & 0.0014 & 0.0011 \\ 
                        & PRL (Full)             & \textbf{0.1003} & \textbf{0.0121 }& \textbf{0.0316} &\textbf{0.0084} & \textbf{0.0064} \\   
\midrule
\multirow{10}{*}{\shortstack{Mexico, Spain, India $\rightarrow$\\ Japan, Germany}} 
                       & CDL (Base Model)               & 0.1127 & 0.0135 & 0.0301 & 0.0086 & 0.0072 \\ 
                        & PRL (Full)             & \textbf{0.1761} & \textbf{0.0230} & \textbf{0.0593} & \textbf{0.0163} & \textbf{0.0123} \\ 
\cmidrule{2-7} 
                       & DLRM (Base Model)               & 0.0756 & 0.0093 & 0.0085 & 0.0041 & 0.0049 \\ 
                        & PRL (Full)             & \textbf{0.2017} &\textbf{ 0.0246} &\textbf{ 0.0545 }&\textbf{ 0.0156} &\textbf{ 0.0131} \\ 
\cmidrule{2-7} 
                       & PerK (Base Model)               & 0.1443 & 0.0177 & 0.0601 & 0.0143 & 0.0094 \\ 
                        & PRL (Full)             & \textbf{0.2750} & \textbf{0.0335} & \textbf{0.1086} & \textbf{0.0263} & \textbf{0.0179} \\ 
\cmidrule{2-7}
                        & NCF (Base Model)               & 0.0096 & 0.0012 &  0.0022 & 0.0007 & 0.0007 \\ 
                        & PRL (Full)             & \textbf{0.1558} & \textbf{0.0202}& \textbf{0.0280} &\textbf{ 0.0107} & \textbf{0.0108} \\    
\cmidrule{2-7}
                        & LightGCN (Base Model)               & 0.0165 & 0.0022 & 0.0061 & 0.0016 & 0.0012 \\ 
                        & PRL (Full)             & \textbf{0.1064} & \textbf{0.0138}& \textbf{0.0278} &\textbf{ 0.0087} & \textbf{0.0077} \\                                             
\bottomrule
\end{tabular}
}
\vskip -0.4 cm
\end{table*}

\begin{table*}[!t]
\centering
\vskip 0.1cm
\caption{Performance of PRL with different base models on MovieLens. The best results are marked with \textbf{bold face}.}
\label{tab:avg_age}
\vskip -0.2cm
\resizebox{0.88\textwidth}{!}{
\setlength{\tabcolsep}{8pt}
\begin{tabular}{c|c|c|c|c|c|c}
\toprule
\textbf{Data} &  \textbf{Method} & \textbf{Recall@20} & \textbf{F1@20} & \textbf{MAP@20} & \textbf{NDCG@20} & \textbf{Precision@20} \\ 
\midrule
\multirow{10}{*}{\shortstack{1, 18, 35, 45, 50, 56 $\rightarrow$\\ 25}}
                        & CDL (Base Model)            & 0.0179 & 0.0274 & 0.0045 & 0.0581 & 0.0587 \\ 
                       & PRL (Full)             & \textbf{0.0252} & \textbf{0.0409} & \textbf{0.0072} & \textbf{0.1071} & \textbf{0.1076} \\ 
\cmidrule{2-7} 
                       & DLRM (Base Model)               & 0.0714 & 0.1096 &\bf 0.0285 & \bf0.2433 & 0.2366 \\ 
                        & PRL (Full)             &\bf{0.0716} & \bf{0.1101} & {0.0284} &{0.2431} & \bf{0.2372} \\ 
\cmidrule{2-7} 
                       & PerK (Base Model)               & 0.0682 & 0.1029 & \textbf{0.0290} & \textbf{0.2224} & 0.2107  \\ 
                        & PRL (Full)             & \textbf{0.0690} & \textbf{0.1037}& 0.0287 &0.2190 & \textbf{0.2110} \\ 
\cmidrule{2-7}
                        & NCF (Base Model)               & 0.0050 & 0.0250 & 0.0011 & 0.0251 & 0.0251 \\ 
                        & PRL (Full)             & \textbf{0.0240} & \textbf{0.0387}& \textbf{0.0057} &\textbf{0.0947} & \textbf{0.1005} \\    
\cmidrule{2-7}
                        & LightGCN (Base Model)               & 0.0081 & 0.0132 & 0.0019 &  0.0381 & 0.0358 \\ 
                        & PRL (Full)             & \textbf{0.0249} & \textbf{0.0402}& \textbf{0.0069} &\textbf{0.1076} & \textbf{0.1055}  \\   
\midrule
\multirow{10}{*}{\shortstack{25 $\rightarrow$\\ 1, 18, 35, 45, 50, 56 }} 
                       & CDL (Base Model)               & 0.0576 & 0.0848 & 0.0174 & 0.1602 & 0.1716 \\ 
                        & PRL (Full)             & \textbf{0.0645} & \textbf{0.0952} & \textbf{0.0202} & \textbf{0.1772} & \textbf{0.1897} \\ 
\cmidrule{2-7} 
                       & DLRM (Base Model)               & 0.0848 & 0.1342 & 0.0382 & 0.3347 & 0.3225 \\ 
                        & PRL (Full)             & \bf{0.0903} &\bf{0.1405} &\bf{0.0414}&\bf{0.3455} &\bf{0.3319} \\ 
\cmidrule{2-7} 
                       & PerK (Base Model)        &  0.0746 & 0.1164 & 0.0324 & 0.2701 & 0.2661       \\ 
                        & PRL (Full)             & \textbf{0.0792} & \textbf{0.1225} & \textbf{0.0355} & \textbf{0.2821} & \textbf{0.2757} \\ 
\cmidrule{2-7}
                        & NCF (Base Model)               & 0.0140 & 0.0229 &  0.0030 & 0.0633 & 0.0652 \\ 
                        & PRL (Full)             & \textbf{0.0450} & \textbf{0.0694}& \textbf{0.0144} &\textbf{ 0.1639} & \textbf{0.1711} \\    
\cmidrule{2-7}
                        & LightGCN (Base Model)               & 0.0093 & 0.0157 & 0.0022 & 0.0497 & 0.0482 \\ 
                        & PRL (Full)             & \textbf{0.0290} & \textbf{0.0480}& \textbf{ 0.0097} &\textbf{0.1493} & \textbf{0.1395} \\                 \bottomrule
\end{tabular}
}
\end{table*}

\section{Experiments}
We evaluate whether PRL improves heterogeneous base recommenders and
whether its causal component benefits cross-domain recommendation.
Additional settings and results can be found in the supplementary material.


\subsection{Datasets}

\textbf{XMRec.} \emph{XMRec}~\cite{xmrec} contains 18 local markets (countries), 16 product categories, and 52.5M user-item interactions. For each item $j$, we use its description as the item feature $\x_j^v$. We exclude users with fewer than three purchases. The three train-test splits are: France, Italy, India $\rightarrow$ Japan, Mexico; Mexico, Spain, India $\rightarrow$ Japan, Germany; and Germany, Italy, Japan $\rightarrow$ United States, India. We use product production countries as the causal confounders $\s_m$.

\textbf{MovieLens.} \emph{MovieLens}~\cite{harper2015movielens} contains movie ratings from users of different ages. We use movie titles and plots as item features $\x_j^v$, and derive user features from the first three films each user rated. We exclude users with fewer than five ratings or with no rating above 3, leaving 6,034 users and 3,705 items. We use two age-based train-test splits: 1--18, 18--25, 35--45, 45--50, 50--56, $56^+$ $\rightarrow$ 25--35; and 25--35 $\rightarrow$ all other groups. For brevity, we denote each age group by its starting age, e.g., ``1'' for ``1--18''. We use normalized movie release years as causal confounders $\s_m$.

In all experiments, we use a cold-start setting where each testing domain user has only one rating in the training set, making the recommendations extremely challenging.

\begin{table*}[!t]
\vskip -0.3cm
\centering
\caption{Comparison between PRL w/o Causality and PRL (Full) on a specific domain with different base models on XMRec. The best results are marked with \textbf{bold face}.}
\vskip -0.3cm
\label{tab:ablation_study_XMRec}
\resizebox{0.88\textwidth}{!}{
\begin{tabular}{c|c|c|c|c|c|c}
\toprule
\textbf{Data} &  \textbf{Method} & \textbf{Recall@20} & \textbf{F1@20} & \textbf{MAP@20} & \textbf{NDCG@20} & \textbf{Precision@20} \\ 
\midrule
\multirow{15}{*}{\shortstack{France, Italy, India $\rightarrow$\\ Japan, Mexico}}
                        & CDL (Base Model)            & 0.0143 & 0.0016 & 0.0028 & 0.0009 & 0.0009 \\ 
                       & CDL PRL w/o Causality      & 0.1058 & 0.0126 & 0.0333 & 0.0088 & 0.0067 \\ 
                       & PRL (Full)             & \textbf{0.1091} & \textbf{0.0128} & \textbf{0.0463} & \textbf{0.0108} & \textbf{0.0068} \\ 
\cmidrule{2-7} 
                       & DLRM (Base Model)               & 0.0044 & 0.0004 & 0.0004 & 0.0002 & 0.0002 \\ 
                        & DLRM PRL w/o Causality      & 0.0232 & 0.0026 & 0.0039 & 0.0014 & 0.0014 \\ 
                        & PRL (Full)             &\textbf{ 0.0295} & \textbf{0.0035} & \textbf{0.0048} &\textbf{ 0.0018} & \textbf{0.0018} \\ 
\cmidrule{2-7} 
                       & PerK (Base Model)               & 0.1098 & 0.0128 & 0.0512 & 0.0112 & 0.0068 \\ 
                       & PerK PRL w/o Causality      & 0.1376 & 0.0160 & 0.0558 & 0.0129 & 0.0085\\ 
                        & PRL (Full)             & \textbf{0.1635} & \textbf{0.0192 }& \textbf{0.0637} &\textbf{ 0.0151} & \textbf{0.0102} \\ 
\cmidrule{2-7}
                        & NCF (Base Model)               & 0.0131 & \textbf{0.0148} & 0.0026 & 0.0008 & 0.0008 \\ 
                       & NCF PRL w/o Causality      & 0.1056 & 0.0126 & 0.0235 & 0.0074 & 0.0067\\ 
                        & PRL (Full)             & \textbf{0.1137} & 0.0137 & \textbf{0.0309} &\textbf{ 0.0090} & \textbf{0.0073} \\    
\cmidrule{2-7}
                        & LightGCN (Base Model)               & 0.0182 & 0.0021 & 0.0050 & 0.0014 & 0.0011 \\ 
                       & LightGCN PRL w/o Causality      & 0.0940 & 0.0112 & 0.0289 & 0.0076 & 0.0059\\ 
                        & PRL (Full)             & \textbf{0.1003} & \textbf{0.0121 }& \textbf{0.0316} &\textbf{0.0084} & \textbf{0.0064} \\ 
                        \bottomrule
\end{tabular}
}
\vskip -0.1 cm
\end{table*}

\begin{figure*}[!tb]
\centering
\vskip -0.1cm
\includegraphics[width=0.9\linewidth]{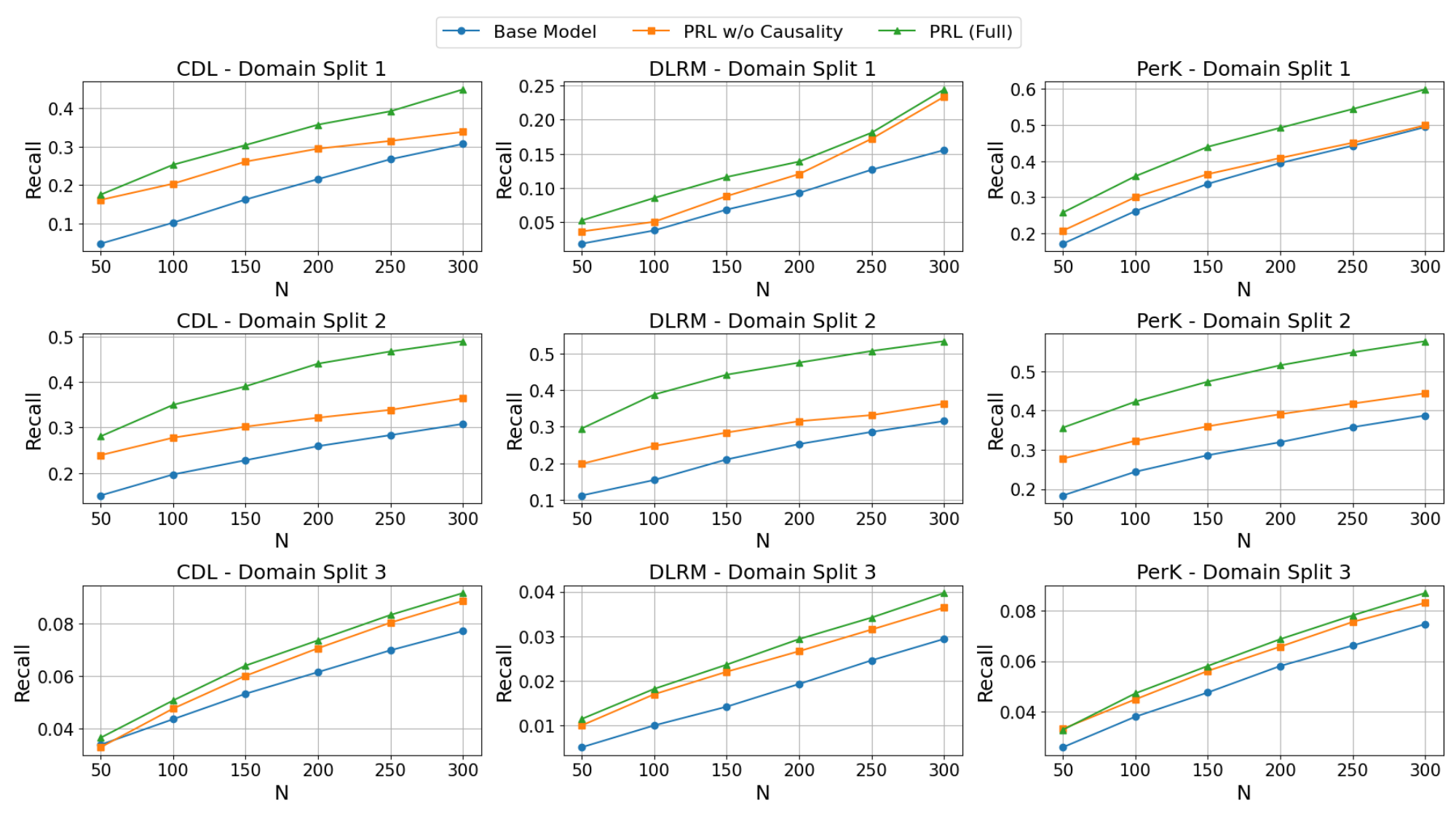}
\vskip -0.5cm
\caption{{Recall@\textit{N} on all three XMRec splits,
comparing each base model with PRL without causality and full PRL.}}
\Description{Recall at different cutoff values on three XMRec
train-test splits. Each plot compares the base recommender, PRL without
causal adjustment, and full PRL for CDL, DLRM, or PerK.}
\label{fig:recall}
\vskip -0.4cm
\end{figure*}

\subsection{Base Recommenders, Baselines, and Metrics}
{PRL is a \emph{plug-and-play} framework compatible with any base recommender. In this paper, we compare PRL with five representative recommendation models, including Collaborative Deep Learning (\textbf{CDL})~\cite{cdl}, Deep Learning Recommendation Model (\textbf{DLRM})~\cite{DLRM}, Top Personalized-K Recommendation. (\textbf{PerK})~\cite{kweon2024perk}, Neural Collaborative Filtering (\textbf{NCF})~\cite{NCF}, and Light Graph Convolutional Network (\textbf{LightGCN})~\cite{lightgcn}. These methods serve as both our \textbf{baselines} and our \textbf{base recommenders} to be enhanced by PRL.}

We evaluate all methods using Recall@20, Precision@20, F1@20, MAP@20, and NDCG@20, averaged over all users; formal metric definitions and training details are in the Appendix.

\subsection{Results}

\textbf{Results for Different Base Models.}
{\tabref{tab:avg_country} and \tabref{tab:avg_age} show that PRL
improves the base models overall. Gains are smaller on MovieLens,
where release year provides a weaker confounding signal than XMRec's
market factors.}

\textbf{Recall@$N$ with Larger $N$.}
{\figref{fig:recall} shows that full PRL consistently
outperforms its non-causal counterpart across base models, splits,
and cutoff values.}

\begin{figure*}[!t]
    \vskip -0.0cm
    \centering
    \includegraphics[width=0.18\textwidth]{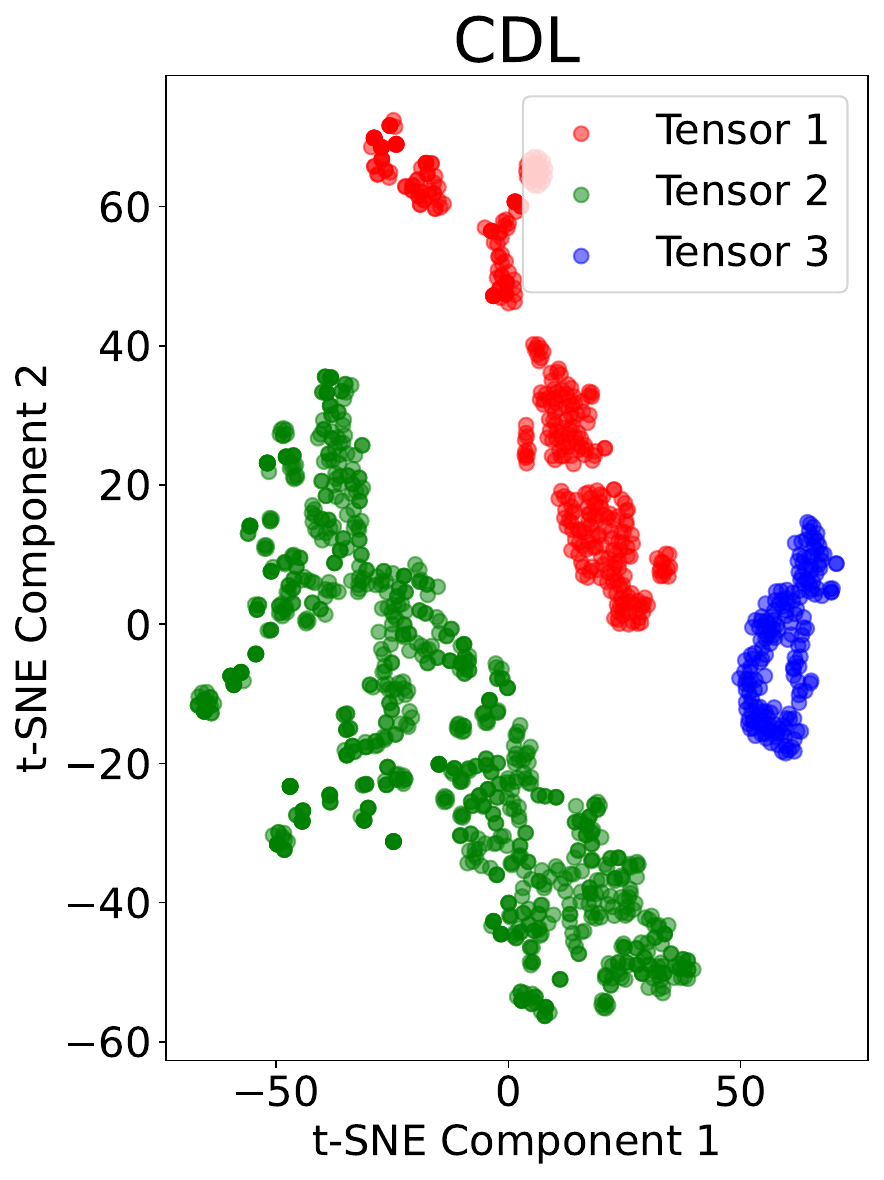}%
    \hfill
    \includegraphics[width=0.18\textwidth]{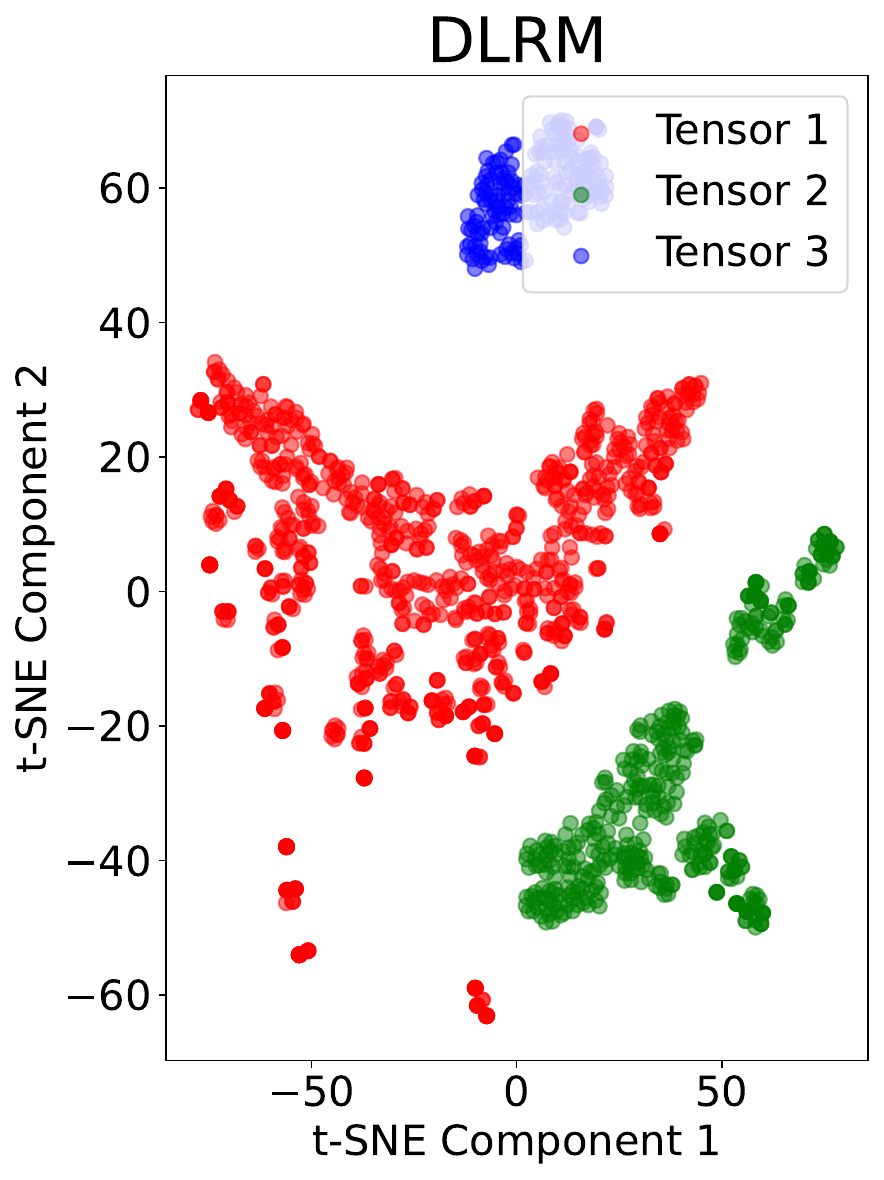}%
    \hfill
    \includegraphics[width=0.18\textwidth]{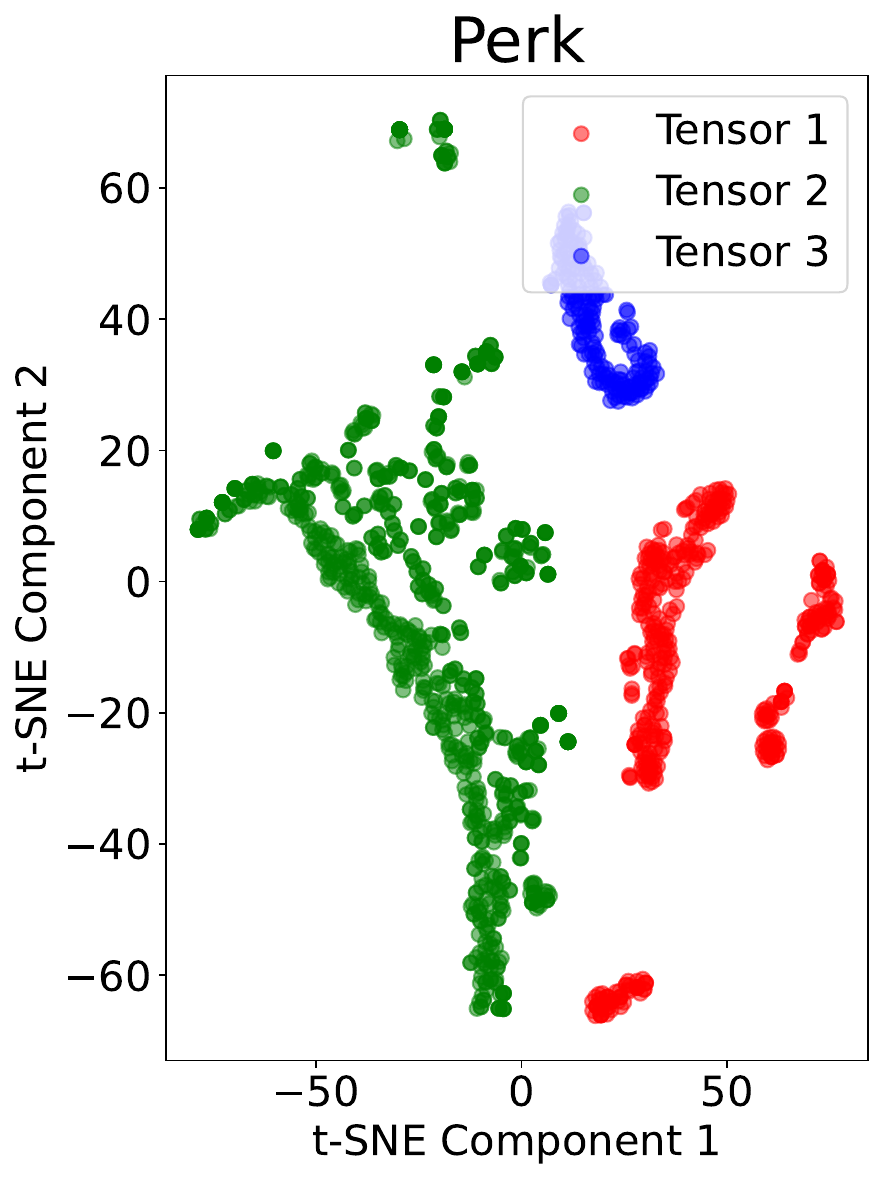}
    \hfill
    \includegraphics[width=0.18\textwidth]{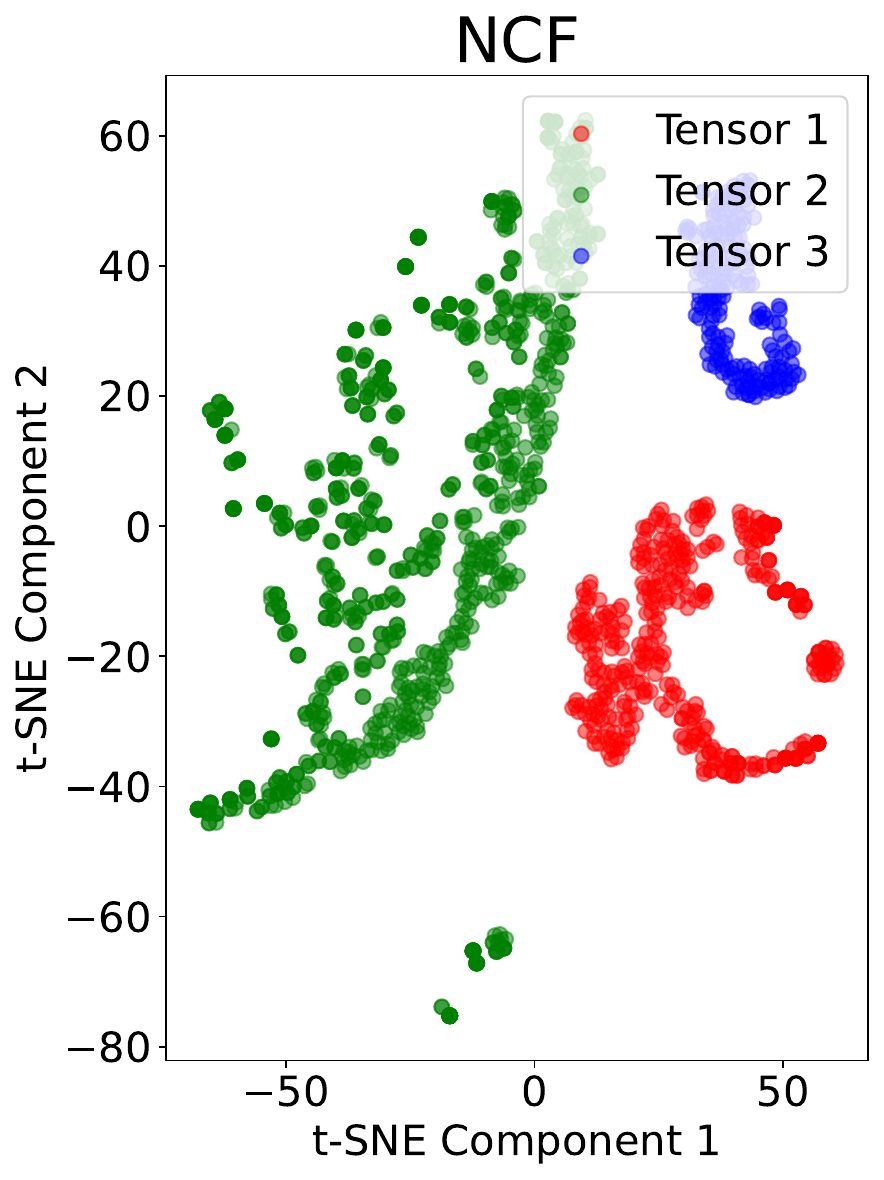}
    \hfill
    \includegraphics[width=0.18\textwidth]{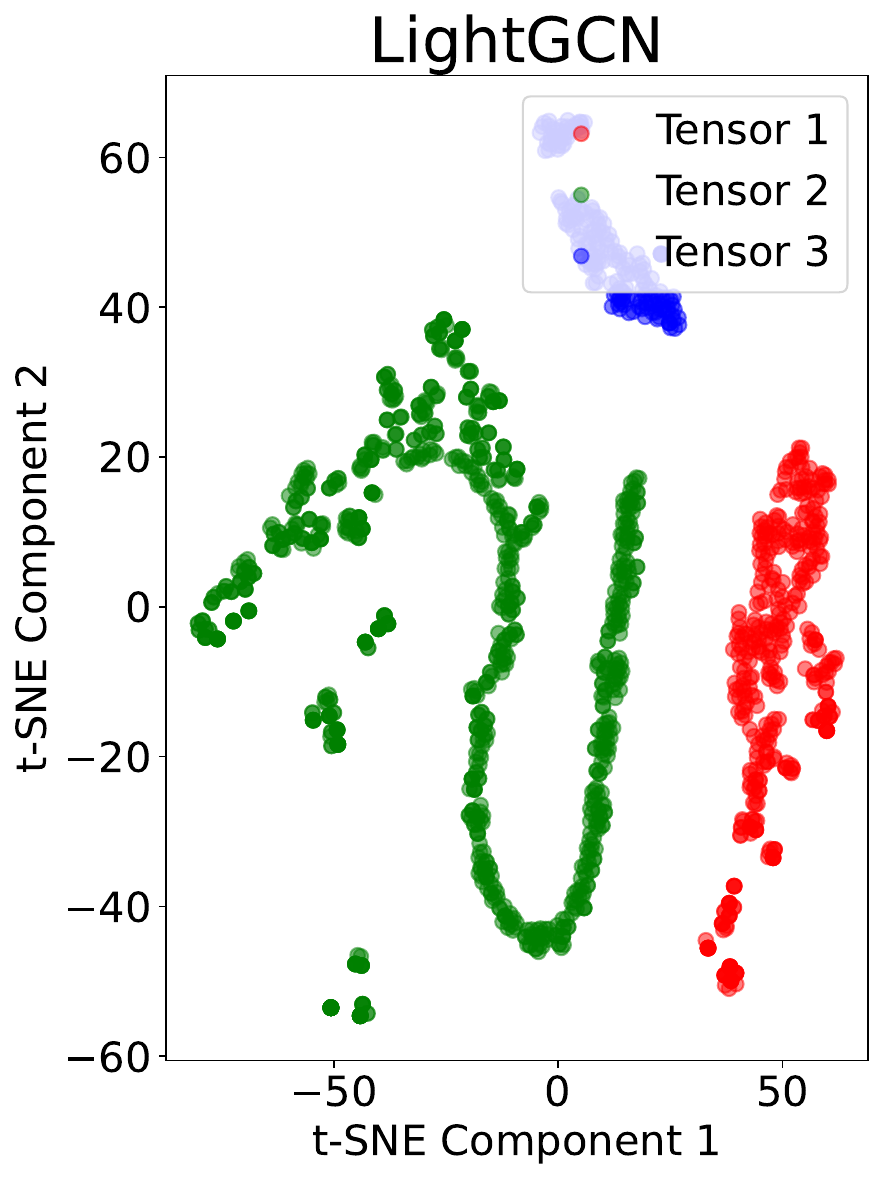}    
    \vskip -0.3 cm
    \caption{{t-SNE projections of PRL user latents $\u_i$ on
    XMRec. Colors denote
    cluster assignments $\pi_i$ inferred by~\eqnref{eq:update_pi}, rather
    than ground-truth labels.}}
    \Description{Five t-SNE projections of the learned user latent
    representations obtained with CDL, DLRM, PerK, NCF, and LightGCN.
    Points are colored according to the three clusters inferred by PRL.}
    \label{fig:visualization}
    \vskip -0.1 cm
\end{figure*}

\begin{figure}[htbp]
    \vskip -0.2cm
    \centering
    \includegraphics[width=1.0\linewidth]{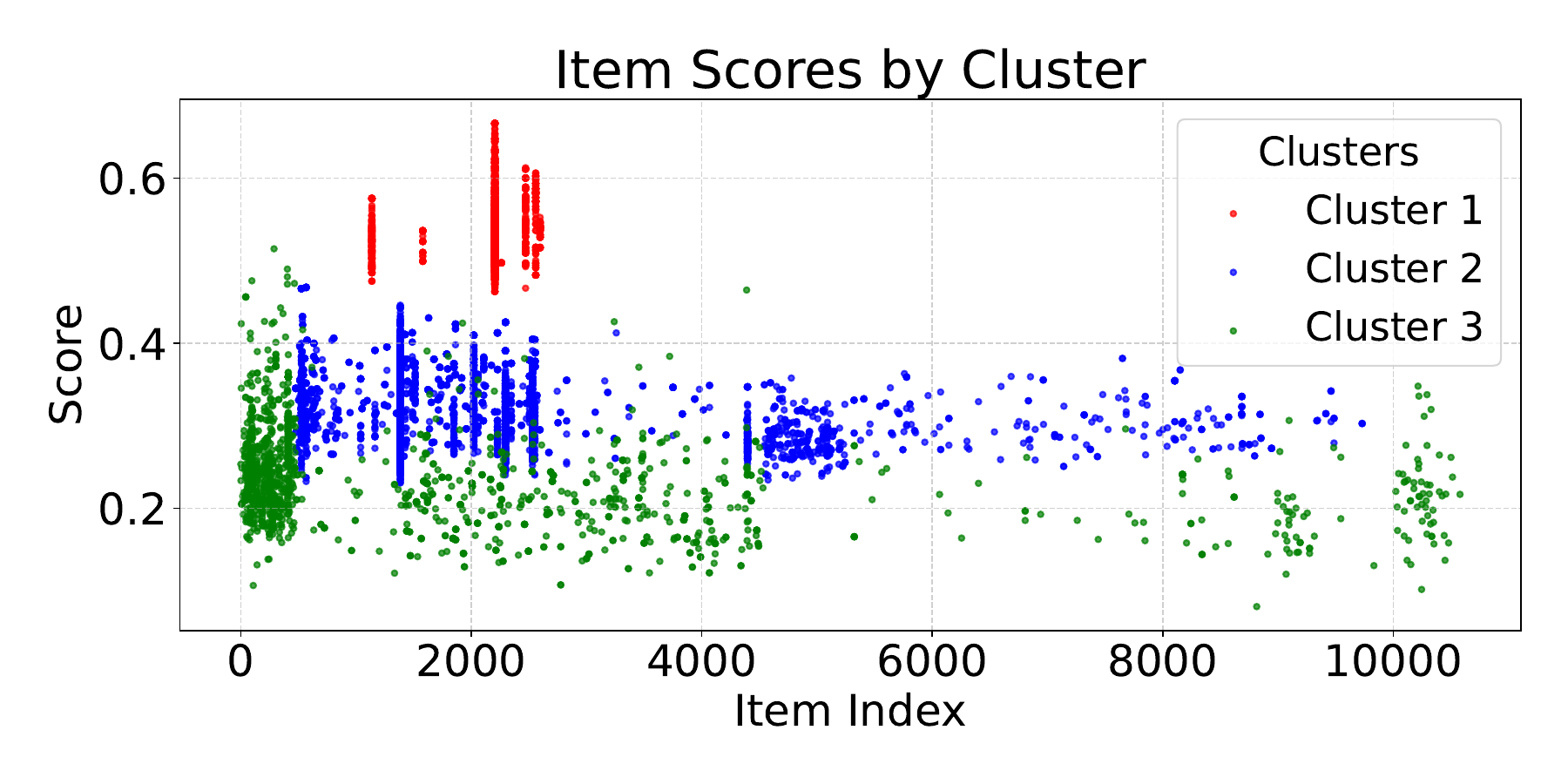}
    \vskip -0.6cm
    \caption{User clusters based on users' highest rated items, using the CDL-based PRL model applied to the XMRec dataset. {X-axis indicates the item ID, while Y-axis indicates the score of the item. Clusters are distinguished by different colors.}}
    \Description{A scatter plot of the highest-rated item selected for each
    user under the CDL-based PRL model. The horizontal axis gives item
    indices and the vertical axis gives rating scores. Points are colored
    according to the three user clusters inferred by PRL, revealing
    cluster-specific concentrations over items and scores.}
    \vskip -0.4cm
    \label{fig:item_score}
\end{figure}

\textbf{Visualizations of the Clusters.} 
\figref{fig:visualization} visualizes the user latent $\u$  for all five base models on the XMRec dataset. Each visualization shows a distinct separation into 3 clusters, indicating successful user grouping of our model. 
Furthermore, Figure~\ref{fig:item_score} illustrates the relationship between user clusters and items using the CDL-based PRL model on the same dataset. {For each user, we selected the item with the highest rating they have given, recorded
the item ID and its rating, and visualized the results. Different clusters are represented using distinct colors, effectively showcasing the distribution and preferences of users within each cluster.} For instance, Cluster 1 (Red) shows pronounced preferences for 4-5 specific items, underscoring the impact of user clustering on improving PRL's performance. 

\textbf{Performance of Each Cluster Discovered by PRL.} For a deeper understanding of the model performance, we include more fine-grained results for different clusters discovered by PRL in Appendix. Results show that our PRL can usually improve performance in most clusters. 

\blue{\textbf{Case Study on Debiasing.} 
To explore PRL's debiasing capability and cross-domain generalization, we conducted a detailed analysis using the first domain pair of the XMRec dataset, which corresponds to the countries \textbf{France, Italy, India, Japan, and Mexico}.}

\blue{We first examined the top-20 recommendations for each user using the CDL baseline model. The resulting distributions showed notable country-specific biases:}

\blue{
\begin{itemize}[nosep,leftmargin=18pt]
  \item Italian users: 34 \textbf{Camera \& Photo} recommendations (3.70\%) out of 920 total recommendations.
  \item Indian users: 878 \textbf{Camera \& Photo} recommendations (6.46\%) out of 13,600 total recommendations.
\end{itemize}}

\blue{The bias ratio across countries was 1.75$\times$ (maximum: 6.46\%, minimum: 3.70\%), indicating that Indian users were recommended camera products 1.75 times more frequently than Italian users.}

\blue{After applying our PRL model, we examined the top-20 recommendations for each user. The resulting distributions were notably more balanced:}
\blue{
\begin{itemize}[nosep,leftmargin=18pt]
  \item Italian users: 33 \textbf{Camera \& Photo} recommendations (3.59\%) out of 920 total recommendations.
  \item Indian users: 427 \textbf{Camera \& Photo} recommendations (3.14\%) out of 13,600 total recommendations.
\end{itemize}}

\blue{The bias ratio across countries was reduced to 1.14$\times$ (maximum: 3.59\%, minimum: 3.14\%), representing a 38.5\% reduction in country-specific bias. This demonstrates that PRL successfully mitigated the preference biases through probabilistic user clustering and causal debiasing, thereby enabling the model to learn more generalizable user-item interaction patterns.}

\textbf{Ablation Study.}
{The comparison in \tabref{tab:ablation_study_XMRec} and \figref{fig:recall} highlights the performance difference between ‘PRL w/o Causality’ and ‘PRL (Full)’. The results consistently show that ‘PRL (Full)’ outperforms its counterpart, ‘PRL w/o Causality’, emphasizing the crucial role of causal inference in enhancing the effectiveness of the PRL model. Furthermore, a comparison between the base model and ‘PRL w/o Causality’ also reveals notable performance improvements, validating the efficacy of PRL's user cluster discovery. Additional details and results can be found in Table S1$\sim$S15 of Appendix.}

\section{Related Work}
\textbf{Domain-Dependent Recommendation.}
Previous work has explored in-domain recommendation scenarios. Early methods, including PMF~\cite{pmf} and BPR~\cite{bpr}, applied collaborative filtering techniques to address {challenges in recommendation}. {Sequential models such as GRU4Rec~\cite{gru4rec},
SASRec~\cite{sasrec}, UniSRec~\cite{unisrec}, and
CauseRec~\cite{causerec} use interaction histories; the latter two
transfer item representations and augment counterfactual sequences,
respectively. CDL~\cite{cdl,CRAE,CKE,ColVAE} instead uses item content
for cold-start recommendation.}

{Despite advances in in-domain recommendation, cross-domain
recommendation remains understudied. Existing work uses domain adaptation
techniques~\cite{VDI,TSDA,UDIL,GRDA,CIDA,DANN}, often relying on
shared users or items across source and target
domains~\cite{DBLP:conf/sigir/Yuan0KZ20,DBLP:conf/sigir/WuYCLH020,
DBLP:conf/sigir/BiSYWWX20a,li2019zero,DBLP:conf/sigir/Hansen0SAL20,
DBLP:conf/sigir/LiangXYY20,DBLP:conf/sigir/ZhuSSC20,
liu2020heterogeneous}. On the other hand, some methods enhance recommendation performance in both source and target domains simultaneously~\cite{0008T20,HuZY18,ZhaoLF19}. In contrast, PRL first infers user clusters and confounders
before cluster-specific recommendation, improving generalization and
robustness to domain shifts.}

\textbf{Causal Inference for Recommendation.}
    Causal inference~\cite{pearl2009causality,CounTS} and discovery~\cite{glymour2019review,ICL} have been widely applied to model cause-and-effect relationships between variables in the machine learning community. Recently, it has been employed to improve the performance of recommender systems~\cite{wang2020causal}. PDA~\cite{zhang2021causal} uses causal intervention to address popularity bias in recommendations, while DICE~\cite{zheng2021disentangling} learns representations from user interactions based on the structured causal model (SCM). {Additionally, some research focuses on debiasing recommendations without adopting a causal inference perspective}~\cite{li2021debiasing, Wang0LZY022,chen2023bias}. {However, these approaches do not consider user groups within the SCM framework.} In contrast, our method divides users into clusters based on a confounder variable and {generates recommendations} by aggregating user ratings through do-calculus, {providing} a more interpretable and sophisticated approach.

\section{Conclusion}
In this paper, we address the problem of cross-domain recommendation by introducing a novel causal Bayesian framework, named {Probabilistic Residual Learning} (PRL). PRL generates recommendations by:  (1) inferring the user cluster ID, (2) inferring the residual rating based on our causal debiasing framework, and (3) predicting the final rating as a correction to the base model's prediction. PRL can enhance the performance of any base recommenders in a plug-and-play manner, and automatically discover meaningful user clusters. As a general probabilistic framework compatible with various recommendation systems, PRL can be extended to additional modalities beyond textual data in future research. 
Furthermore, PRL provides interpretability by uncovering latent user preferences and biases that influence rating predictions. Its modular design also allows seamless integration with deep
learning-based recommenders, making it a scalable and adaptable
solution for diverse recommendation scenarios.

\bibliographystyle{ACM-Reference-Format}
\bibliography{main}

\clearpage

\twocolumn[
\begin{center}
    {\LARGE\bfseries 
    Probabilistic Residual Learning for Online Recommendations (Appendix)}
    \vspace{0.5em}
\end{center}
\vspace{1em}
]

\appendix
\begin{bibunit}[ACM-Reference-Format]
\setcounter{table}{0}
\renewcommand{\thetable}{S\arabic{table}}

\setcounter{figure}{0}
\renewcommand{\thefigure}{S\arabic{figure}}

\setcounter{equation}{0}
\renewcommand{\theequation}{S\arabic{equation}}

\section{More Details on Experiments and Implementation}

\subsection{Metrics}
\label{app:metrics}

\textbf{mAP.} mAP is defined as:
\begin{align}
    \text{AP}_i = \frac{1}{|J_i|} \sum_{n=1}^{N} \text{rel}_{i,n} \times \text{Precision}_i@n,
\end{align}
where $N$ is the total number of recommended items, $\text{Precision}_i@n$ is the precision at rank $n$, and $|J_i|$ is the total number of relevant items for user $i$. The mean Average Precision (mAP) is then calculated by averaging $\text{AP}_i$ over all users:
\begin{align}
    \text{mAP} = \frac{1}{|I|} \sum_{i=1}^{|I|} \text{AP}_i,
\end{align}
where $|I|$ is the total number of users.

\textbf{NDCG.} NDCG@\textit{N} is computed as follows.

First, the Discounted Cumulative Gain (DCG@\textit{N}) is calculated:
\begin{align}
\text{DCG}_i@N = \sum_{n=1}^{N} \frac{2^{\text{rel}_{i,n}} - 1}{\log_2(n + 1)},
\end{align}
where \( \text{rel}_{i,n} \) denotes the relevance of the item at position \( n \) for user \( i \).
Next, the Ideal Discounted Cumulative Gain (IDCG@\textit{N}), representing the maximum possible DCG (i.e., all relevant items ranked at the top), is calculated as:
\begin{align}
\text{IDCG}_i@N = \sum_{n=1}^{\min(N, |J_i|)} \frac{2^{1} - 1}{\log_2(n + 1)} = \sum_{n=1}^{\min(N, |J_i|)} \frac{1}{\log_2(n + 1)},
\end{align}
where \( |J_i| \) denotes the total number of relevant items for user \( i \).

Finally, the Normalized Discounted Cumulative Gain is obtained by normalizing DCG@\textit{N} by IDCG@\textit{N}:
\begin{align}
\text{NDCG}_i@N = \frac{\text{DCG}_i@N}{\text{IDCG}_i@N}.
\end{align}

Here the logarithmic term \( \log_2(n + 1) \) discounts the relevance based on the item's position in the ranked list, serving as the normalization factor.

\subsection{Training Configurations}\label{app:train_config}
Following CDL~\cite{cdl}, we set the hidden dimension $h=50$ for all latent vectors, as well as for the encoder network. Similar to CDL, we also add an additional decoder to reconstruct $\x^v$, serving as a regularization term during training. 
During training, we use AdamW~\cite{Adam} as our optimizer, with a learning rate of $10^{-3}$ and a batch size of $256$. The base models were trained for 100 epochs, while PRL was trained for 150 epochs. All experiments were conducted on an NVIDIA RTX A5000 GPU.

\subsection{Performance of Each Clusters Discovered by PRL}\label{app:fine_grained_experiments}

\tabref{tab:cluster_cdl}, \ref{tab:cluster_dlrm}, \ref{tab:cluster_perk}, \ref{tab:cluster_NCF}, \ref{tab:cluster_LightGCN} show PRL's performance across different clusters on XMRec using CDL, DLRM, PerK, NCF, and LightGCN as base models. These results support the conclusion that PRL improves upon the base models even without incorporating the causality component. Furthermore, the full PRL consistently outperforms its non-causal counterpart across all configurations. For example, CDL, as the base model, achieves a recall@$20$ of $0.0241$ for User Cluster 1 in the split of ``France, Italy, India $\rightarrow$ Japan, Mexico''. When PRL without the causal inference component is applied, recall improves to $0.0278$. The full PRL further enhances performance for this metric, achieving a recall@$20$ of $0.0708$.

{\tabref{tab:cluster_cdl_movie}, \ref{tab:cluster_DLRM_movie}, \ref{tab:cluster_perk_movie}, \ref{tab:cluster_NCF_movie}, \ref{tab:cluster_LightGCN_movie} show PRL's performance across different clusters on MovieLens with the same five base models. Even with some fluctuations, the similar improvements are consistent with the results for XMRec.}


\subsection{Ablation Study}
\label{app:ablation_experiments}

The performance comparison across \tabref{tab:cluster_cdl_movie}-\ref{tab:cluster_LightGCN} shows that ``PRL (Full)'' {generally} outperforms ``PRL w/o Causality'', highlighting the effectiveness of causal inference in PRL. Additionally, comparing the base model with ``PRL w/o Causality'' reveals performance enhancements, suggesting that PRL's user cluster discovery significantly boosts performance. 

\blue{\textbf{Simple Baseline.} Moreover, we conducted experiments by clustering the user features and then performing per-cluster modeling to predict the residuals.  \tabref{no_reg} presents the results comparing this simple baseline with our full PRL, verifying that our latent variable modeling is highly effective, while per-cluster modeling alone yields limited improvements.}

\blue{\textbf{Larger Base Models.} We also conducted additional ablation experiments by scaling up the CDL baseline to approximately match the parameter size of the PRL-enriched network. Concretely, we expanded both the depth and width of the CDL architecture as suggested. The original CDL structure was \textbf{512 → 200 → 50}, while the larger (deeper and wider) version used an architecture of \textbf{512 → 550 → 400 → 50}, resulting in roughly \textbf{1.05M} parameters, which is comparable to PRL’s \textbf{0.90M} parameters. This ensures that the comparison isolates the effect of PRL from mere model capacity differences. The results are presented in \tabref{larger_CDL}. Although the deeper CDL exhibits a slight performance improvement over the base model, PRL still achieves substantially higher performance across most metrics. This demonstrates that PRL’s gains stem from its ability to capture cross-domain relational patterns, user-cluster-specific representations, rather than simply from an increased parameter count.}

\blue{\textbf{More User Records in the Training Set.} We conducted experiments using CDL as the base model to analyze the effect of incorporating a higher number of testing-domain user records into the training process. Specifically, we used the second domain pair of the XMRec dataset, involving users from Mexico, Spain, India, Japan, and Germany. we selected users with more than four interactions and varied the number of testing-domain user records included in the training set to construct 1-shot, 2-shot, and 3-shot scenarios, where $n$-shot means there are $n$ records for each testing user in the training set. The performance of both the base CDL model and our proposed PRL model in these settings is summarized in \tabref{tab:cdl_nshot} and \tabref{tab:pruc_nshot} below. }

\begin{table*}[!t]
\centering
\caption{Performance of PRL on different user clusters with CDL as the base model on MovieLens. ``-'' means a cluster contains only training-set users, i.e., no test-set users to evaluate. The best results are marked with \textbf{bold face}.}
\label{tab:cluster_cdl_movie}
\vskip -0.2cm
\resizebox{0.9\textwidth}{!}{
\begin{tabular}{c|c|c|c|c|c|c|c}
\toprule
\textbf{Data} & \textbf{Cluster} & \textbf{Method} & \textbf{Recall@20} & \textbf{F1@20} & \textbf{MAP@20} & \textbf{NDCG@20} & \textbf{Precision@20} \\ 
\midrule
\multirow{9}{*}{\shortstack{1, 18, 35, 45, 50, 56 $\rightarrow$\\ 25}}
                      &   & CDL (Base Model)              & 0.0 & 0.0 & 0.0 & 0.0 & 0.0 \\ 
                      & 1 & PRL w/o Causality      & 0.0 & 0.0 & 0.0 & 0.0 & 0.0 \\ 
                      &   & PRL (Full)            & 0.0 & 0.0 & 0.0 & 0.0 & 0.0 \\ 
\cmidrule{2-8} 
                      &  & CDL (Base Model)              & - & - & - & - & - \\ 
                      & 2  & PRL w/o Causality      & - & - & - & - & - \\ 
                      &   & PRL (Full)             & - & - & - & - & -  \\ 
\cmidrule{2-8} 
                      &  & CDL (Base Model)              & 0.0179 & 0.0274 & 0.0045 & 0.0581 & 0.0587 \\ 
                      & 3 & PRL w/o Causality     &  0.0186 & 0.0302 & 0.0056 & 0.0864 & 0.0802 \\ 
                      &   & PRL (Full)             & \bf0.0252 & \bf0.0409 & \bf0.0072 & \bf0.1071 & \bf0.1077 \\ 
\midrule
\multirow{9}{*}{\shortstack{25 $\rightarrow$\\ 1, 18, 35, 45, 50, 56}}
                      &   & CDL (Base Model)             & \bf0.0558 & \bf0.0861 & \bf0.0174 & 0.1758 & \bf0.1879 \\ 
                      & 1 & PRL w/o Causality      & {0.0317} & 0.0528 & 0.0095 &  0.1511 & 0.1572 \\ 
                      &   & PRL (Full)             & \bf0.0558 & \bf0.0861 & \bf0.0174 & \bf0.1759 & \bf0.1879 \\ 
\cmidrule{2-8} 
                      &   & CDL (Base Model)              & 0.0651 & 0.0795 & 0.0173 & 0.0938 & 0.1020 \\ 
                      & 2  & PRL w/o Causality      & 0.0676 & 0.0880 & 0.0183 & 0.1159 & 0.1259 \\ 
                      &   & PRL (Full)             & \textbf{0.1016} & \textbf{0.1341} & \textbf{0.0319} & \textbf{0.1832} & \textbf{0.1972} \\ 
\cmidrule{2-8} 
                      &  & CDL (Base Model)              & - & - & - & - & - \\ 
                      & 3  & PRL w/o Causality      & - & - & - & - & - \\ 
                      &   & PRL (Full)             & - & - & - & - & - \\ 
\bottomrule
\end{tabular}
}
\end{table*}

\begin{table*}[!t]
\centering
\caption{Performance of PRL with different base models on MovieLens. The best results are marked with \textbf{bold face}.}
\vskip -0.2cm
\label{tab:ablation_study_MovieLens}
\resizebox{1.0\textwidth}{!}{
\setlength{\tabcolsep}{6pt}
\begin{tabular}{c|c|c|c|c|c|c}
\toprule
\textbf{Data} &  \textbf{Method} & \textbf{Recall@20} & \textbf{F1@20} & \textbf{MAP@20} & \textbf{NDCG@20} & \textbf{Precision@20} \\ 
\midrule
\multirow{15}{*}{\shortstack{1, 18, 35, 45, 50, 56 $\rightarrow$\\ 25}}
                        & CDL (Base Model)            & 0.0179 & 0.0274 & 0.0045 & 0.0581 & 0.0587 \\ 
                       & PRL w/o Causality      &  0.0186 & 0.0302 & 0.0057 & 0.0863 & 0.0801 \\ 
                       & PRL (Full)             & \textbf{0.0252} & \textbf{0.0409} & \textbf{0.0072} & \textbf{0.1071} & \textbf{0.1076} \\ 
\cmidrule{2-7} 
                       & DLRM (Base Model)               & 0.0714 & 0.1096 &\bf 0.0285 & \bf0.2433 & 0.2366 \\ 
                        & PRL w/o Causality      & 0.0232 & 0.0026 & 0.0039 & 0.0014 & 0.0014 \\ 
                        & PRL (Full)             &\bf{0.0716} & \bf{0.1101} & {0.0284} &{0.2431} & \bf{0.2372} \\ 
\cmidrule{2-7} 
                       & PerK (Base Model)               & 0.0682 & 0.1029 & \textbf{0.0290} & \textbf{0.2224} & 0.2107  \\ 
                       & PRL w/o Causality      & 0.0582 & 0.0877 & 0.0212 & 0.1755 & 0.1787\\ 
                        & PRL (Full)             & \textbf{0.0690} & \textbf{0.1037}& 0.0287 &0.2190 & \textbf{0.2110} \\ 
\cmidrule{2-7}
                        & NCF (Base Model)               & 0.0050 & 0.0250 & 0.0011 & 0.0251 & 0.0251 \\ 
                       & PRL w/o Causality      & 0.0231 & 0.0374 & 0.0055 & 0.0927 & 0.0989\\ 
                        & PRL (Full)             & \textbf{0.0240} & \textbf{0.0387}& \textbf{0.0057} &\textbf{0.0947} & \textbf{0.1005} \\    
\cmidrule{2-7}
                        & LightGCN (Base Model)               & 0.0081 & 0.0132 & 0.0019 &  0.0381 & 0.0358 \\ 
                       & PRL w/o Causality      & 0.0248 & 0.0402 & 0.0070 & 0.1077 & 0.1053\\ 
                        & PRL (Full)             & \textbf{0.0249} & \textbf{0.0402}& \textbf{0.0069} &\textbf{0.1076} & \textbf{0.1055}  \\   
\bottomrule
\end{tabular}
}
\vskip -0.5 cm
\end{table*}

\begin{table*}[!t]
\centering
\caption{Performance of PRL on different user clusters with DLRM as the base model on MovieLens. ``-'' means a cluster contains only training-set users, i.e., no test-set users to evaluate. The best results are marked with \textbf{bold face}.}
\label{tab:cluster_DLRM_movie}
\vskip -0.2cm
\resizebox{0.9\textwidth}{!}{
\begin{tabular}{c|c|c|c|c|c|c|c}
\toprule
\textbf{Data} & \textbf{Cluster} & \textbf{Method} & \textbf{Recall@20} & \textbf{F1@20} & \textbf{MAP@20} & \textbf{NDCG@20} & \textbf{Precision@20} \\ 
\midrule
\multirow{9}{*}{\shortstack{1, 18, 35, 45, 50, 56 $\rightarrow$\\ 25}}
                      &   & DLRM (Base Model)              - & - & - & - & - \\ 
                      & 1 & PRL w/o Causality      - & - & - & - & - \\ 
                      &   & PRL (Full)             - & - & - & - & - \\ 
\cmidrule{2-8} 
                      &  & DLRM (Base Model)              & 0.0714 & 0.1097 & \bf0.0285 & \bf0.2433 & 0.2367 \\ 
                      & 2  & PRL w/o Causality      & 0.0269 & 0.0434 & 0.0073 & 0.1078 & 0.1112 \\ 
                      &   & PRL (Full)             & \bf0.0716 & \bf0.1101 & 0.0284 &0.2431 & \bf0.2372  \\ 
\cmidrule{2-8} 
                      &  & DLRM (Base Model)              & 0.0 & 0.0 & 0.0 & 0.0 & 0.0\\ 
                      & 3 & PRL w/o Causality     & 0.0 & 0.0 & 0.0 & 0.0 & 0.0 \\ 
                      &   & PRL (Full)             & 0.0 & 0.0 & 0.0 & 0.0 & 0.0 \\ 
\midrule
\multirow{9}{*}{\shortstack{25 $\rightarrow$\\ 1, 18, 35, 45, 50, 56}}
                      &   & DLRM (Base Model)             & 0.0790 & 0.1264 & 0.0343 & 0.3266 & 0.3146 \\ 
                      & 1 & PRL w/o Causality      & {0.0328} & 0.0548 & 0.0116 &  0.1716 & 0.1656 \\ 
                      &   & PRL (Full)             & \bf0.0848 & \bf0.1366 & \bf0.0396 & \bf0.3634 & \bf0.3505 \\ 
\cmidrule{2-8} 
                      &   & DLRM (Base Model)              & 0.0882 & 0.1390 & 0.0405 & \bf0.3396 & \bf0.3271 \\ 
                      & 2  & PRL w/o Causality      & 0.0975 & 0.1382 & 0.0426 & 0.2572 & 0.2374 \\ 
                      &   & PRL (Full)             & \textbf{0.1119} & \textbf{0.1561} & \bf{0.0486} & {0.2745} & {0.2583} \\ 
\cmidrule{2-8} 
                      &  & DLRM (Base Model)              & - & - & - & - & - \\ 
                      & 3  & PRL w/o Causality      & - & - & - & - & - \\ 
                      &   & PRL (Full)             & - & - & - & - & - \\ 
\bottomrule
\end{tabular}
}
\end{table*}

\begin{table*}[!t]
\centering
\caption{Performance of PRL on different user clusters with Perk as the base model on MovieLens. ``-'' means a cluster contains only training-set users, i.e., no test-set users to evaluate. The best results are marked with \textbf{bold face}.}
\label{tab:cluster_perk_movie}
\vskip -0.2cm
\resizebox{\textwidth}{!}{
\begin{tabular}{c|c|c|c|c|c|c|c}
\toprule
\textbf{Data} & \textbf{Cluster} & \textbf{Method} & \textbf{Recall@20} & \textbf{F1@20} & \textbf{MAP@20} & \textbf{NDCG@20} & \textbf{Precision@20} \\ 
\midrule
\multirow{9}{*}{\shortstack{1, 18, 35, 45, 50, 56 $\rightarrow$\\ 25}}
                      &  & Perk (Base Model)             & - & - & - & - & - \\ 
                      & 1  & PRL w/o Causality      & - & - & - & - & -  \\ 
                      &   & PRL (Full)            & - & - & - & - & -\\ 
\cmidrule{2-8} 
                      &  & Perk (Base Model)              & \bf0.0686 & \bf0.1040 & \bf0.0295 & \bf0.2271 & \bf0.2150 \\ 
                      & 2  & PRL w/o Causality      &  0.0583 & 0.0884 & 0.0215 &  0.1788 & 0.1820\\ 
                      &   & PRL (Full)             & 0.0683 & 0.1036 & 0.0288 & 0.2215 & 0.2136  \\ 
\cmidrule{2-8} 
                      &  &Perk (Base Model)              & 0.0585 & 0.0745 & 0.0173 & 0.1053 & 0.1023 \\ 
                      & 3 & PRL w/o Causality     &  0.0550 & 0.0701 & 0.0139 &  0.0942 & 0.0967 \\ 
                      &   & PRL (Full)             & \bf0.0847 & \bf0.1074 & \bf0.0277& \bf0.1559 & \bf0.1467 \\ 
\midrule
\multirow{9}{*}{\shortstack{25 $\rightarrow$\\ 1, 18, 35, 45, 50, 56}}
                      &  & Perk (Base Model)             & \bf0.0745 & \bf0.1179 &\bf0.0332 & 0.2868 & 0.2826 \\ 
                      & 1  & PRL w/o Causality      & {0.0319} & {0.0530} & {0.0140} & { 0.1811} & {0.1563} \\ 
                      &   & PRL (Full)             & \bf0.0745 & \bf0.1179 &\bf 0.0332 & \bf0.2870 &\bf 0.2828 \\ 
\cmidrule{2-8} 
                      &  & Perk (Base Model)              & 0.0750 & 0.1090 & 0.0292 & 0.2033 & 0.1995 \\ 
                      & 2  & PRL w/o Causality       &  0.0939 & 0.1338 & 0.0367 & 0.2399 & 0.2323 \\ 
                      &   & PRL (Full)            & \bf0.0984 &\bf 0.1407 & \bf0.0446 & \bf0.2628 & \bf0.2469\\ 
\cmidrule{2-8} 
                      &  & Perk (Base Model)              & - & - & - & - & - \\ 
                      & 3  & PRL w/o Causality       & - & - & - & - & - \\ 
                      &   & PRL (Full)             & - & - & - & - & -\\ 
\bottomrule
\end{tabular}
}
\end{table*}

\begin{table*}[!t]
\centering
\caption{Performance of PRL on different user clusters with NCF as the base model on MovieLens. ``-'' means a cluster contains only training-set users, i.e., no test-set users to evaluate. The best results are marked with \textbf{bold face}.}
\label{tab:cluster_NCF_movie}
\vskip -0.2cm
\resizebox{\textwidth}{!}{
\begin{tabular}{c|c|c|c|c|c|c|c}
\toprule
\textbf{Data} & \textbf{Cluster} & \textbf{Method} & \textbf{Recall@20} & \textbf{F1@20} & \textbf{MAP@20} & \textbf{NDCG@20} & \textbf{Precision@20} \\ 
\midrule
\multirow{9}{*}{\shortstack{1, 18, 35, 45, 50, 56 $\rightarrow$\\ 25}}
                      &   & NCF (Base Model)              & 0.0 & 0.0 & 0.0 & 0.0 & 0.0 \\ 
                      & 1 & PRL w/o Causality      & 0.0 & 0.0 & 0.0 & 0.0 & 0.0  \\ 
                      &   & PRL (Full)            & \bf0.1964 & \bf0.1325 & \bf0.0230 & \bf0.0754 & \bf0.1000 \\ 
\cmidrule{2-8} 
                      &  & NCF (Base Model)              & 0.0051 & 0.0087 & 0.0011 & 0.0282 & 0.0279 \\ 
                      & 2  & PRL w/o Causality      &  0.0271 & 0.0443 & 0.0067 &  0.1134 & 0.1210 \\ 
                      &   & PRL (Full)             & \bf0.0285 & \bf0.0463 & \bf0.0070 & \bf0.1159 & \bf0.1231  \\ 
\cmidrule{2-8} 
                      &  & NCF (Base Model)              & 0.0047 & 0.0074 & 0.0009 & 0.0172 & 0.0177 \\ 
                      & 3 & PRL w/o Causality     &  \bf0.0125 & \bf0.0192 & \bf0.0023 & 0.0386 & 0.0409 \\ 
                      &   & PRL (Full)             & 0.0120 & 0.0185 & 0.0022 & \bf0.0389 & \bf0.0410 \\ 
\midrule
\multirow{9}{*}{\shortstack{25 $\rightarrow$\\ 1, 18, 35, 45, 50, 56}}
                      &  & NCF (Base Model)             & 0.0149 & 0.0248 & 0.0032 & 0.0710 & 0.0729 \\ 
                      & 1  & PRL w/o Causality      & \textbf{0.0309} & \textbf{0.0515} & \textbf{0.0088} & \textbf{ 0.1494} & \textbf{0.1555} \\ 
                      &   & PRL (Full)             & 0.0306 & 0.0512 & 0.0087 & 0.1484 & 0.1551 \\ 
\cmidrule{2-8} 
                      &  & NCF (Base Model)              & - & - & - & - & - \\ 
                      & 2  & PRL w/o Causality       & - & - & - & - & - \\ 
                      &   & PRL (Full)            & - & - & - & - & -\\ 
\cmidrule{2-8} 
                      &  & NCF (Base Model)              & 0.0098 & 0.0150 & 0.0021 & 0.0302 & 0.0319 \\ 
                      & 3  & PRL w/o Causality      & 0.0941 & 0.1316 & 0.0312 & 0.2094 & 0.2185 \\ 
                      &   & PRL (Full)             & \bf0.1071 & \bf0.1481 & \bf0.0392 & \bf0.2309 &\bf 0.2402 \\ 
\bottomrule
\end{tabular}
}
\end{table*}

\begin{table*}[!t]
\centering
\caption{Performance of PRL on different user clusters with LightGCN as the base model on MovieLens. ``-'' means a cluster contains only training-set users, i.e., no test-set users to evaluate. The best results are marked with \textbf{bold face}.}
\label{tab:cluster_LightGCN_movie}
\vskip -0.2cm
\resizebox{\textwidth}{!}{
\begin{tabular}{c|c|c|c|c|c|c|c}
\toprule
\textbf{Data} & \textbf{Cluster} & \textbf{Method} & \textbf{Recall@20} & \textbf{F1@20} & \textbf{MAP@20} & \textbf{NDCG@20} & \textbf{Precision@20} \\ 
\midrule
\multirow{9}{*}{\shortstack{1, 18, 35, 45, 50, 56 $\rightarrow$\\ 25}}
                      &   & LightGCN (Base Model)              & - & - & - & - & - \\ 
                      & 1 & PRL w/o Causality     & - & - & - & - & -  \\ 
                      &   & PRL (Full)            & - & - & - & - & - \\ 
\cmidrule{2-8} 
                      &  & LightGCN (Base Model)              & 0.0081 & 0.0132 & 0.0019 & 0.0381 & 0.0358 \\ 
                      & 2  & PRL w/o Causality      &  \bf0.0248 & \bf0.0402 & \bf0.0070 & \bf0.1075 & 0.1052 \\ 
                      &   & PRL (Full)             & \bf0.0248 & 0.0401 & 0.0069 & 0.1073 & \bf0.1053  \\ 
\cmidrule{2-8} 
                      &  & LightGCN (Base Model)              & 0.0226 & 0.0224 & 0.0075 & 0.0227 & 0.0222 \\ 
                      & 3 & PRL w/o Causality     &  0.0214 & 0.0378 & 0.0115 & 0.1911 & 0.1611 \\ 
                      &   & PRL (Full)             & \bf0.0563 & \bf0.0884 & \bf0.0226 & \bf0.2219 & \bf0.2056 \\ 
\midrule
\multirow{9}{*}{\shortstack{25 $\rightarrow$\\ 1, 18, 35, 45, 50, 56}}
                      &  & LightGCN (Base Model)             & 0.0094 & 0.0157 & 0.0022 & 0.0498 & 0.0484 \\ 
                      & 1  & PRL w/o Causality      & \textbf{0.0300} & \textbf{0.0495} & \textbf{0.0101} & \textbf{ 0.1515} & \textbf{0.1416} \\ 
                      &   & PRL (Full)             & 0.0288 & 0.0477 & 0.0097 & 0.1492 & 0.1394 \\ 
\cmidrule{2-8} 
                      &  & LightGCN (Base Model)              & 0.0068 & 0.0110 & 0.0011 & 0.0277 & 0.0294 \\ 
                      &  2 & PRL w/o Causality      & 0.0297 & 0.0428 & 0.0130 & 0.0953 & 0.0765 \\ 
                      &   & PRL (Full)             & \textbf{0.0531} & \textbf{0.0793} & \textbf{0.0204} & \textbf{0.1597} & \textbf{0.1559} \\ 
\cmidrule{2-8} 
                      &  & LightGCN (Base Model)              & - & - & - & - & - \\ 
                      &  3 & PRL w/o Causality      & - & - & - & - & - \\ 
                      &   & PRL (Full)             & - & - & - & - & - \\ 
\bottomrule
\end{tabular}
}
\end{table*}


\begin{table*}[!t]
\small
\centering
\caption{Performance of PRL on different user clusters with CDL as the base model on XMRec. ``-'' means a cluster contains only training-set users, i.e., no test-set users to evaluate. The best results are marked with \textbf{bold face}.}
\label{tab:cluster_cdl}
\vskip -0.2cm
\resizebox{\textwidth}{!}{
\begin{tabular}{c|c|c|c|c|c|c|c}
\toprule
\textbf{Data} & \textbf{Cluster} & \textbf{Method} & \textbf{Recall@20} & \textbf{F1@20} & \textbf{MAP@20} & \textbf{NDCG@20} & \textbf{Precision@20} \\ 
\midrule
\multirow{9}{*}{\shortstack{France, Italy, India $\rightarrow$\\ Japan, Mexico}}
                      &   & CDL (Base Model)              & 0.0241 & 0.0028 & 0.0062 & 0.0018 & 0.0015 \\ 
                      & 1 & PRL w/o Causality      & \bf0.1972 & \bf0.0238 & \bf0.0905 & \bf0.0197 & \bf0.0127 \\ 
                      &   & PRL (Full)             & {0.0708} & {0.0074} & {0.0652} & {0.0105} & {0.0039} \\ 
\cmidrule{2-8} 
                      &  & CDL (Base Model)              & 0.0126 & 0.0014 & 0.0022 & 0.0007 & 0.0008 \\ 
                      & 2  & PRL w/o Causality      & 0.0902 & 0.0107 & 0.0236 & 0.0069 & 0.0057 \\ 
                      &   & PRL (Full)             & \textbf{0.1156} & \textbf{0.0138} & \textbf{0.0431} & \textbf{0.0109} & \textbf{0.0073} \\ 
\cmidrule{2-8} 
                      &  & CDL (Base Model)              & - & - & - & - & - \\ 
                      & 3 & PRL w/o Causality     & - & - & - & - & - \\ 
                      &   & PRL (Full)             & - & - & - & - & - \\ 
\midrule
\multirow{9}{*}{\shortstack{Mexico, Spain, India $\rightarrow$\\ Japan, Germany}}
                      &  & CDL (Base Model)             & 0.1742 & 0.0225 & 0.0333 & 0.0123 & 0.0120 \\ 
                      & 1  & PRL w/o Causality      & \textbf{0.2114} & \textbf{0.0267} & \textbf{0.0707} & \textbf{0.0194} & \textbf{0.0142} \\ 
                      &   & PRL (Full)             & 0.1665 & 0.0222 & 0.0634 & 0.0170 & 0.0119 \\ 
\cmidrule{2-8} 
                      &  & CDL (Base Model)              & 0.0903 & 0.0102 & 0.0289 & 0.0072 & 0.0054 \\ 
                      & 2  & PRL w/o Causality      & 0.1532 & 0.0187 & 0.0524 & 0.0136 & 0.0100 \\ 
                      &   & PRL (Full)             & \textbf{0.1796} & \textbf{0.0233} & \textbf{0.0579} & \textbf{0.0160} & \textbf{0.0124} \\ 
\cmidrule{2-8} 
                      &  & CDL (Base Model)              & - & - & - & - & - \\ 
                      & 3  & PRL w/o Causality      & - & - & - & - & - \\ 
                      &   & PRL (Full)             & - & - & - & - & - \\ 
\midrule
\multirow{9}{*}{\shortstack{Germany, Italy, Japan $\rightarrow$\\ United States, India}}
                      &  & CDL (Base Model)             & 0.0262 & 0.0059 & \textbf{0.0079} & 0.0041 & 0.0033 \\ 
                      & 1  & PRL w/o Causality      & {0.0261} & {0.0063} &  0.0072 & \textbf{0.0044} & {0.0036} \\ 
                      &   & PRL (Full)             & \textbf{0.0266} & \textbf{0.0064} & 0.0062 & 0.0042 & \textbf{0.0037} \\ 
\cmidrule{2-8} 
                      &  & CDL (Base Model)              & 0.0244 & {0.0054} & \textbf{0.0088} & \textbf{0.0042} & \textbf{0.0031} \\ 
                      & 2  & PRL w/o Causality      & 0.0166 & 0.0037 & 0.0041 & 0.00234 & 0.0021 \\ 
                      &   & PRL (Full)             & \textbf{0.0250} &\textbf {0.0055} & \textbf{0.0088} & \textbf{0.0042} & \textbf{0.0031} \\ 
\cmidrule{2-8} 
                      &  & CDL (Base Model)              & {0.0277} & \textbf{0.0049} & {0.0066} & {0.0028} & \textbf{0.0027} \\ 
                      &  3 & PRL w/o Causality      & 0.0194 & 0.0045 & 0.0049 &\bf0.0030 & 0.0026 \\ 
                      &   & PRL (Full)             & \textbf{0.0278} & \textbf{0.0049} & \textbf{0.0067} & {0.0028} & \textbf{0.0027} \\ 
\bottomrule
\end{tabular}
}
\end{table*}


\begin{table*}[h]
\caption{Performance of PRL on different user clusters with DLRM as the base model on XMRec. ``-'' means a cluster contains only training-set users, i.e., no test-set users to evaluate. The best results are marked with \textbf{bold face}.}
\label{tab:cluster_dlrm}
\vskip -0.1 cm
\resizebox{\textwidth}{!}{
\begin{tabular}{c|c|c|c|c|c|c|c}
\toprule
\textbf{Data} & \textbf{Cluster} & \textbf{Method} & \textbf{Recall@20} & \textbf{F1@20} & \textbf{MAP@20} & \textbf{NDCG@20} & \textbf{Precision@20} \\ 
\midrule
\multirow{9}{*}{\shortstack{France, Italy, India $\rightarrow$\\ Japan, Mexico}}
                      &  & DLRM (Base Model)              & 0.0051 & 0.0005 & 0.0004 & 0.0002 & 0.0003 \\ 
                      & 1  & PRL w/o Causality      & 0.0246 & 0.0027 & 0.0039 & 0.0014 &  0.0014 \\ 
                      &   & PRL (Full)             & \bf{0.0345} & \bf{0.004} & \bf{0.0056} & \bf{0.0021} & \bf{0.0021} \\ 
\cmidrule{2-8} 
                      &  & DLRM (Base Model)              & 0.0000 & 0.0000 & 0.0000 & 0.0000 & 0.0000 \\ 
                      &  2 & PRL w/o Causality      & \bf{0.0150} & \bf{0.0017} & \bf{0.0040} & \bf{0.0010} & \bf{0.0009} \\ 
                      &   & PRL (Full)             & 0.0000 & 0.0000 & 0.0000 & 0.0000 & 0.0000 \\ 
\cmidrule{2-8} 
                      &  & DLRM (Base Model)              & - & - & - & - & - \\ 
                      & 3 & PRL w/o Causality      & - & - & - & - & -\\ 
                      &   & PRL (Full)             & - & - & - & - & - \\ 
\midrule
\multirow{9}{*}{\shortstack{Mexico, Spain, India $\rightarrow$\\ Japan, Germany}}
                      &  & DLRM (Base Model)             & 0.0000 & 0.0000 & 0.0000 & 0.0000 & 0.0000 \\ 
                      &  1 & PRL w/o Causality      & \bf0.3296 & \bf0.0416 & {0.0203} & \bf0.0153 & \bf0.0222 \\ 
                      &   & PRL (Full)             & {0.3074} & {0.0395} & \bf0.0213 & {0.0152} & {0.0211} \\ 
\cmidrule{2-8} 
                      &  & DLRM (Base Model)              & 0.0780 & 0.0096 & 0.0087 & 0.0042 & 0.0051 \\ 
                      &  2 & PRL w/o Causality      & 0.1398 & 0.0174 & 0.0277 & 0.0096 & 0.0093 \\ 
                      &   & PRL (Full)             & \textbf{0.1984} & \textbf{0.0241} & \textbf{0.0555} & \textbf{0.0157} & \textbf{0.0128} \\ 
\cmidrule{2-8} 
                      & & DLRM (Base Model)              & - & - & - & - & - \\ 
                      & 3   & PRL w/o Causality      & - & - & - & - & - \\ 
                      &   & PRL (Full)             & - & - & - & - & - \\ 
\midrule
\multirow{9}{*}{\shortstack{Germany, Italy, Japan $\rightarrow$\\ United States, India}}
                      & & DLRM (Base Model)             & 0.0023 & 0.0006 & 0.0003 & 0.0003 & 0.0003 \\ 
                      & 1   & PRL w/o Causality      & 0.0042 & \bf0.0011 & \textbf{0.0010} & \textbf{0.0007 }& \textbf{0.0007} \\ 
                      &   & PRL (Full)             & \textbf{0.0046} & \textbf{0.0011} & 0.0009 & \bf0.0007 & {0.0006} \\ 
\cmidrule{2-8} 
                      &  & DLRM (Base Model)              & 0.0018 & 0.0005 & 0.0003 & 0.0003 & 0.0003 \\ 
                      & 2  & PRL w/o Causality      & \textbf{0.0045} & \textbf{0.0012} & 0.0010 & \textbf{0.0007} &  \textbf{0.0007} \\ 
                      &   & PRL (Full)             & \bf0.0045 & 0.0011 & \textbf{0.0012} & \bf0.0007 & \bf0.0007 \\ 
\cmidrule{2-8} 
                      &  & DLRM (Base Model)              & 0.0036 & 0.0008 & 0.0005 & 0.0004 & 0.0004 \\ 
                      &  3 & PRL w/o Causality      & 0.0052 & 0.0015 & 0.0009 & 0.0009 & 0.0009 \\ 
                      &   & PRL (Full)             & \textbf{0.0141} & \textbf{0.0034} & \textbf{0.0075} & \textbf{0.0032} & \textbf{0.0019} \\ 
\bottomrule
\end{tabular}
}
\end{table*}

\begin{table*}[h]
\centering
\small
\caption{Performance of PRL on different user clusters with PerK as the base model on XMRec. ``-'' means a cluster contains only training-set users, i.e., no test-set users to evaluate. The best results are marked with \textbf{bold face}.}
\vskip -0.17 cm
\label{tab:cluster_perk}
\resizebox{\textwidth}{!}{
\begin{tabular}{c|c|c|c|c|c|c|c}
\toprule
\textbf{Data} & \textbf{Cluster} & \textbf{Method} & \textbf{Recall@20} & \textbf{F1@20} & \textbf{MAP@20} & \textbf{NDCG@20} & \textbf{Precision@20} \\ 
\midrule
\multirow{9}{*}{\shortstack{France, Italy, India $\rightarrow$\\ Japan, Mexico}}
                      &   & PerK (Base Model)              & {0.1752} & {0.0204} & {0.1152} & {0.022} & {0.0108} \\ 
                      & 1 & PRL w/o Causality      & \bf0.2153 &\bf 0.0260 &\bf 0.1255 & \bf0.0252 & \bf0.0139 \\ 
                      &   & PRL (Full)             & {0.1782} & {0.0210} & {0.1162} & {0.0226} & {0.0114} \\ 
\cmidrule{2-8} 
                      &  & PerK (Base Model)             & 0.0986 & 0.0115 & 0.0403 & 0.0094 & 0.0061 \\ 
                      & 2  & PRL w/o Causality      & {0.1243} & {0.0143} & {0.0440} & {0.0108} & {0.0076}  \\ 
                      &   & PRL (Full)             & \bf0.1629 & \bf0.0189 & \bf0.0548 & \bf0.0138 & \bf0.0100 \\ 
\cmidrule{2-8} 
                      &  & PerK (Base Model)              & - & - & - & - & - \\ 
                      & 3 & PRL w/o Causality      & - & - & - & - & - \\ 
                      &   & PRL (Full)             & - & - & - & - & - \\ 
\midrule
\multirow{9}{*}{\shortstack{Mexico, Spain, India $\rightarrow$\\ Japan, Germany}}
                      &  & PerK (Base Model)             & 0.1434 & 0.0176 & 0.0582 & 0.014 & 0.0094 \\ 
                      & 1  & PRL w/o Causality      & 0.2175 & 0.0262 & 0.0913 & 0.0217 &  0.0140 \\ 
                      &   & PRL (Full)             & \bf{0.2905} & \bf{0.0353} & \bf{0.1157} & \bf{0.0278} & \bf{0.0188} \\ 
\cmidrule{2-8} 
                      &  & PerK (Base Model)              & 0.1495 & 0.0184 & 0.0723 & 0.0166 & 0.0098\\ 
                      & 2  & PRL w/o Causality      & \bf{0.2783} & \bf{0.0307} & \bf{0.0964} & \bf{0.0232} &  \bf{0.0163} \\ 
                      &   & PRL (Full)             & 0.1790 &  0.0224 & 0.0646 & 0.0167 & 0.0120 \\ 
\cmidrule{2-8} 
                      &  & PerK (Base Model)              & - & - & - & - & - \\ 
                      & 3  & PRL w/o Causality      & - & - & - & - & -\\ 
                      &   & PRL (Full)             & - & - & - & - & - \\ 
\midrule
\multirow{9}{*}{\shortstack{Germany, Italy, Japan $\rightarrow$\\ United States, India}}
                      &  & PerK (Base Model)             & 0.0194 & 0.0043 & 0.0057 & 0.003 & 0.0024 \\ 
                      &  1 & PRL w/o Causality      & 0.0295 & {0.0066} & \bf{0.0087} &  \bf{0.0046} & {0.0037} \\ 
                      &   & PRL (Full)             & \bf{0.0308} & \bf0.0068 & 0.0086 & \bf0.0046 &\bf 0.0038 \\ 
\cmidrule{2-8} 
                      &  & PerK (Base Model)              & 0.0126 & 0.0028 & 0.0032 & 0.0018 & 0.0016 \\ 
                      & 2  & PRL w/o Causality      & {0.0155} & 0.0035 & 0.0040 & {0.0022} & {0.0020} \\ 
                      &   & PRL (Full)            & \bf0.0162 & \bf0.0037 & \bf{0.0048} & \bf{0.0025} & \bf0.0021 \\ 
\cmidrule{2-8} 
                      &  & PerK (Base Model)              & {0.0261} & {0.0035} & {0.0091} & {0.0025} & {0.0019} \\ 
                      & 3  & PRL w/o Causality      & 0.0174 & 0.0027 & 0.0013 & 0.0012 & 0.0014 \\ 
                      &   & PRL (Full)             & \bf{0.0266} & \bf{0.0041} & \bf{0.0102} & \bf{0.0033} & \bf{0.0022} \\ 
\bottomrule
\end{tabular}
}
\end{table*}

\begin{table*}[h]
\centering
\vskip -0.1 cm
\caption{Performance of PRL on different user clusters with NCF as the base model on XMRec. ``-'' means a cluster contains only training-set users, i.e., no test-set users to evaluate. The best results are marked with \textbf{bold face}.}
\vskip -0.17 cm
\label{tab:cluster_NCF}
\resizebox{\textwidth}{!}{
\begin{tabular}{c|c|c|c|c|c|c|c}
\toprule
\textbf{Data} & \textbf{Cluster} & \textbf{Method} & \textbf{Recall@20} & \textbf{F1@20} & \textbf{MAP@20} & \textbf{NDCG@20} & \textbf{Precision@20} \\ 
\midrule
\multirow{9}{*}{\shortstack{France, Italy, India $\rightarrow$\\ Japan, Mexico}}
                      &   & NCF (Base Model)            &  0.0090	& 0.0010 &	0.0019 &	0.0005	 & 0.0005\\ 
                      & 1 & PRL w/o Causality      & \bf0.2013 & \bf0.0238 & \bf0.0537 & \bf0.0151 &  \bf0.0127 \\ 
                      &   & PRL (Full)             & 	{0.1581} &	{0.0176}	  & {0.0476}	 & {0.0122}	 & {0.0093}\\ 
\cmidrule{2-8} 
                      &  & NCF (Base Model)             &0.0165	&0.0019	&0.0032	&0.0010	&0.0010 \\ 
                      & 2  & PRL w/o Causality      & {0.0893} & {0.0107} & {0.0184} & {0.0061} & {0.0057}  \\ 
                      &   & PRL (Full)            & \bf0.1062 &	\bf0.0130&	\bf0.0280&	\bf0.0084	&\bf0.0069 \\ 
\cmidrule{2-8} 
                      &  & NCF (Base Model)              & - & - & - & - & - \\ 
                      & 3 & PRL w/o Causality      & - & - & - & - & - \\ 
                      &   & PRL (Full)             & - & - & - & - & - \\ 
\midrule
\multirow{9}{*}{\shortstack{Mexico, Spain, India $\rightarrow$\\ Japan, Germany}}
                      &  & NCF (Base Model)             & 0.0097 &  0.0013 & 0.0022 & 0.0007 &0.0007 \\ 
                      & 1  & PRL w/o Causality      & 0.1081 & 0.0142 & 0.0181 & 0.0073 &  0.0076 \\ 
                      &   & PRL (Full)             & \bf{0.1560} & \bf{0.0202} & \bf{0.0280} & \bf{0.0107} & \bf{0.0108} \\ 
\cmidrule{2-8} 
                      &  & NCF (Base Model)              & - & - & - & - & -\\ 
                      & 2  & PRL w/o Causality       & - & - & - & - & - \\ 
                      &   & PRL (Full)             & - & - & - & - & - \\ 
\cmidrule{2-8} 
                      &  & NCF (Base Model)             & 0.0 & 0.0 & 0.0 & 0.0 & 0.0\\ 
                      & 3  & PRL w/o Causality      & 0.0 & 0.0 & 0.0 & 0.0 & 0.0 \\ 
                      &   & PRL (Full)             & 0.0 & 0.0 & 0.0 & 0.0 & 0.0 \\ 
\midrule
\multirow{9}{*}{\shortstack{Germany, Italy, Japan $\rightarrow$\\ United States, India}}
                      &  & NCF (Base Model)             & 0.0020 & 0.0005 &0.0006 & 0.0004 & 0.0003 \\ 
                      & 1  & PRL w/o Causality      & 0.0204 & {0.0051} & \bf{0.0041} &  {0.0030} & {0.0029} \\ 
                      &   & PRL (Full)            & \bf0.0214 &\bf 0.0055 & {0.0039} & \bf{0.0032} & \bf0.0031 \\ 
\cmidrule{2-8} 
                      &  & NCF (Base Model)              & 0.0018 & 0.0005 & 0.0003 &  0.0003 &  0.0003 \\ 
                      & 2  & PRL w/o Causality      & {0.0064} & 0.0015 & 0.0008 & {0.0008} & {0.0009} \\ 
                      &   & PRL (Full)            & \bf0.0079 &  \bf{0.0021} & \bf{0.0011} & \bf0.0011 & \bf0.0012 \\ 
\cmidrule{2-8} 
                      &  & NCF (Base Model)              & 0.0 & 0.0 & 0.0 & 0.0 & 0.0 \\ 
                      & 3  & PRL w/o Causality      & 0.0 & 0.0 & 0.0 & 0.0 & 0.0 \\ 
                      &   & PRL (Full)             & 0.0 & 0.0 & 0.0 & 0.0 & 0.0\\ 
\bottomrule
\end{tabular}
}
\end{table*}

\begin{table*}[h]
\centering
\vskip -0.1 cm
\caption{Performance of PRL on different user clusters with LightGCN as the base model on XMRec. ``-'' means a cluster contains only training-set users, i.e., no test-set users to evaluate. The best results are marked with \textbf{bold face}.}
\vskip -0.17 cm
\label{tab:cluster_LightGCN}
\resizebox{\textwidth}{!}{
\begin{tabular}{c|c|c|c|c|c|c|c}
\toprule
\textbf{Data} & \textbf{Cluster} & \textbf{Method} & \textbf{Recall@20} & \textbf{F1@20} & \textbf{MAP@20} & \textbf{NDCG@20} & \textbf{Precision@20} \\ 
\midrule
\multirow{9}{*}{\shortstack{France, Italy, India $\rightarrow$\\ Japan, Mexico}}
                      &   & LightGCN (Base Model)              &	0.0261	&0.0034	&0.0028& 0.0015	& 0.0018 \\ 
                      & 1 & PRL w/o Causality      & \bf0.1742 & \bf0.0209 & \bf0.0749 & \bf0.0167 & \bf0.0111 \\ 
                      &   & PRL (Full)            & 0.1400	&0.0154	&0.0482	&0.0115	&0.0081\\ 
\cmidrule{2-8} 
                      &  & LightGCN (Base Model)            & 0.0168	&0.0019	&0.0054	&0.0013 &0.0010\\ 
                      & 2  & PRL w/o Causality      & {0.0804} & {0.0095} & {0.0211} & {0.0060} & {0.0051}  \\ 
                      &   & PRL (Full)            &\bf0.0936	&\bf0.0115	&\bf0.0288&	\bf0.0079	&\bf0.0061\\ 
\cmidrule{2-8} 
                      &  & LightGCN (Base Model)     & - & - & - & - & -\\ 
                      & 3 & PRL w/o Causality      & - & - & - & - & - \\ 
                      &   & PRL (Full)             & - & - & - & - & - \\ 
\midrule
\multirow{9}{*}{\shortstack{Mexico, Spain, India $\rightarrow$\\ Japan, Germany}}
                      &  & LightGCN (Base Model)             & 0.0093 & 0.0009 & 0.0046 &0.0008 & 0.0005 \\ 
                      & 1  & PRL w/o Causality      & \bf0.0972 & \bf0.0097 & \bf0.0129 & \bf0.0045 & \bf 0.0051 \\ 
                      &   & PRL (Full)             & 0.0 & 0.0 & 0.0 & 0.0 & 0.0 \\ 
\cmidrule{2-8} 
                      &  & LightGCN (Base Model)              & 0.0170 & 0.0023 & 0.0062 & 0.0017 &  0.0012 \\ 
                      & 2  & PRL w/o Causality      & {0.1040} & {0.0135} & {0.0215} & {0.0077} &  {0.0072} \\ 
                      &   & PRL (Full)             &\bf 0.1790 & \bf 0.0224 & \bf0.0646 & \bf0.0167 & \bf0.0120 \\ 
\cmidrule{2-8} 
                      &  & LightGCN (Base Model)              & - & - & - & - & - \\ 
                      & 3  & PRL w/o Causality      & - & - & - & - & -\\ 
                      &   & PRL (Full)             & - & - & - & - & - \\ 
\midrule
\multirow{9}{*}{\shortstack{Germany, Italy, Japan $\rightarrow$\\ United States, India}}
                      &  & LightGCN (Base Model)             & 0.0016 & 0.0005 & 0.0002 & 0.0002 & 0.0003 \\ 
                      & 1  & PRL w/o Causality      & 0.0062 & \bf{0.0017} & {0.0012} &  {0.0010} & \bf{0.0010} \\ 
                      &   & PRL (Full)             & \bf{0.0066} & \bf0.0017 & \bf0.0014 &\bf0.0011 & \bf0.0010 \\ 
\cmidrule{2-8} 
                      &  & LightGCN (Base Model)             & 0.0 & 0.0 & 0.0 & 0.0 & 0.0 \\ 
                      &  2 & PRL w/o Causality      & 0.0 & 0.0 & 0.0 & 0.0 & 0.0 \\ 
                      &   & PRL (Full)             & 0.0 & 0.0 & 0.0 & 0.0 & 0.0\\ 
\cmidrule{2-8} 
                      &  & LightGCN (Base Model)              & { 0.0016} & {0.0004} & {.0002} & {0.0002} & {0.0002} \\ 
                      & 3  & PRL w/o Causality      & 0.0037 & \bf0.0009 & 0.0010 & \bf0.0006 & \bf0.0005 \\ 
                      &   & PRL (Full)             & \bf{0.0039} & {0.0008} & \bf{0.0012} & \bf{0.0006} & \bf{0.0005} \\ 
\bottomrule
\end{tabular}
}
\vskip -0.3cm
\end{table*}

\blue{
\begin{table*}[!t]
\vskip 0.3cm
\centering
\caption{\blue{Results on the simple baseline and PRL.}}
\label{no_reg}
\resizebox{1.0\textwidth}{!}{
\begin{tabular}{c|c|c|c|c|c}
\toprule
\textbf{Method} & \textbf{Recall@20} & \textbf{F1@20} & \textbf{MAP@20} & \textbf{NDCG@20} & \textbf{Precision@20} \\
\midrule
                         CDL           & 0.0143 & 0.0016 & 0.0028 & 0.0009 & 0.0009 \\ 
                        Clustering + Residuals      & 0.0156 & 0.0018 & 0.0023 & 0.0009 & 0.0010 \\ 
                        PRL (Full)             & \textbf{0.1091} & \textbf{0.0128} & \textbf{0.0463} & \textbf{0.0108} & \textbf{0.0068} \\ 
\midrule
                         DLRM               & 0.0044 & 0.0004 & 0.0004 & 0.0002 & 0.0002 \\ 
                         Clustering + Residuals      & 0.0163 & 0.0018 & 0.0029 & 0.0009 & 0.0010 \\ 
                         PRL (Full)             & \textbf{0.0295} & \textbf{0.0035} & \textbf{0.0048} & \textbf{0.0018} & \textbf{0.0018} \\ 
\midrule
                         PerK               & 0.1098 & 0.0128 & 0.0512 & 0.0112 & 0.0068 \\ 
                         Clustering + Residuals      & 0.1118 & 0.0129 & 0.0513 & 0.0113 & 0.0069 \\ 
                         PRL (Full)             & \textbf{0.1635} & \textbf{0.0192} & \textbf{0.0637} & \textbf{0.0151} & \textbf{0.0102} \\ 
\midrule
                         NCF              & 0.0131 & 0.00148 & 0.0026 & 0.0008 & 0.0008 \\ 
                         Clustering + Residuals      & 0.0164 & 0.0019 & 0.0029 & 0.0010 & 0.0010 \\ 
                         PRL (Full)             & \textbf{0.1137} & \textbf{0.0137} & \textbf{0.0309} & \textbf{0.0090} & \textbf{0.0073} \\    
\midrule
                         LightGCN             & 0.0182 & 0.0021 & 0.0050 & 0.0014 & 0.0011 \\ 
                         Clustering + Residuals      & 0.0277 & 0.0031 & 0.0054 & 0.0017 & 0.0017 \\ 
                         PRL (Full)             & \textbf{0.1003} & \textbf{0.0121} & \textbf{0.0316} & \textbf{0.0084} & \textbf{0.0064} \\ 
\bottomrule
\end{tabular}
}
\vskip -0.3 cm
\end{table*}
}

\blue{
\begin{table*}[!t]
\vskip 0.3cm
\centering
\caption{\blue{Performance comparison between larger CDL and PRL.}}
\label{larger_CDL}
\resizebox{1.0\textwidth}{!}{
\begin{tabular}{c|c|c|c|c|c}
\toprule
\textbf{Method} & \textbf{Recall@20} & \textbf{F1@20} & \textbf{MAP@20} & \textbf{NDCG@20} & \textbf{Precision@20} \\
\midrule
                         CDL           & 0.0143 & 0.0016 & 0.0028 & 0.0009 & 0.0009 \\ 
                       CDL (Larger)    &   0.0223 & 0.0026 & 0.0022 & 0.0011 & 0.0014  \\ 
                        PRL (Full)             & \textbf{0.1091} & \textbf{0.0128} & \textbf{0.0463} & \textbf{0.0108} & \textbf{0.0068} \\ 
\bottomrule
\end{tabular}
}
\vskip -0.3 cm
\end{table*}
}

\blue{
\begin{table*}[!t]
\vskip 0.3cm
\centering
\caption{\blue{Results in $n$-shot settings for the CDL base model.}}
\label{tab:cdl_nshot}
\resizebox{1.0\textwidth}{!}{
\begin{tabular}{c|c|c|c|c|c|c}
\toprule
\textbf{Method} & \textbf{Config} & \textbf{Recall@20} & \textbf{Precision@20} & \textbf{F1@20} & \textbf{MAP@20} & \textbf{NDCG@20} \\
\midrule
CDL (Base Model) & 1-shot & 0.0647 & \textbf{0.0105} & 0.0181 & 0.0111 & 0.0098 \\
CDL (Base Model) & 2-shot & 0.0700 & 0.0085 & \textbf{0.0152} & 0.0108 & 0.0078 \\
CDL (Base Model) & 3-shot & \textbf{0.0817} & 0.0060 & 0.0112 & \textbf{0.0159} & \textbf{0.0106} \\
\bottomrule
\end{tabular}
}
\vskip -0.3cm
\end{table*}
}

\begin{table*}[!t]
\vskip 0.3cm
\captionof{table}{\blue{Results in $n$-shot settings for our proposed model PRL (Full) using CDL as the base model.}}
\label{tab:pruc_nshot}
\resizebox{1.0\textwidth}{!}{
\begin{tabular}{c|c|c|c|c|c|c}
\toprule
\textbf{Method} & \textbf{Config} & \textbf{Recall@20} & \textbf{Precision@20} & \textbf{F1@20} & \textbf{MAP@20} & \textbf{NDCG@20} \\
\midrule
PRL (Full) & 1-shot & 0.1080 & \textbf{0.0195} & \textbf{0.0359} & 0.0252 & 0.0208 \\
PRL (Full) & 2-shot & 0.1507 & \textbf{0.0195} & 0.0356 & 0.0359 & 0.0216 \\
PRL (Full) & 3-shot & \textbf{0.2178} & 0.0145 & 0.0272 & \textbf{0.0471} & \textbf{0.0251} \\
\bottomrule
\end{tabular}
}
\end{table*}

\clearpage

\noindent
\begin{minipage}[t]{\columnwidth}
\renewcommand{\refname}{References}
\putbib[main]
\end{minipage}

\end{bibunit}

\end{document}